\documentclass[prd,aps,twocolumn,nofootinbib]{revtex4-1}

\usepackage{graphicx}
\usepackage{epsfig}

\usepackage{amssymb}
\usepackage{amsmath}

\usepackage{url}
\usepackage[bookmarks, pagebackref=true]{hyperref}
\usepackage{color}
	\definecolor{rossoCP3}{cmyk}{0,.88,.77,.40}
		\definecolor{graa}{rgb}{0.8,0.8,0.8}
		\definecolor{blaa}{rgb}{0.2,0.2,0.6}
		\definecolor{gron}{RGB}{0,150,0}
		\hypersetup{
			colorlinks, 
			bookmarksopen, 
			bookmarksnumbered,
			citecolor=blaa, 		
			linkcolor= rossoCP3,	
			urlcolor=blaa,			
			}
\usepackage[margin=5pt, font=small,labelfont=bf,justification=raggedright]{caption}
\usepackage{subfig}

\newcommand{\ea}[1]{
\begin{align}
#1
\end{align}
}

\newcommand{\sea}[1]{
\begin{subequations}
\begin{align}
#1
\end{align}
\end{subequations}
}

\newcommand{\seal}[2]{
\begin{subequations}
\label{#1}
\begin{align}
#2
\end{align}
\end{subequations}
}

\newcommand{\nn}{\nonumber \\ }
\newcommand{\MSB}{{\overline{\rm MS}}}
\newcommand{\R}{{\cal R}}
\newcommand{\s}{{\cal S}}
\newcommand{\rd}{{\rm d}}

\newcommand{\tgamma}{\tilde{\gamma}}
\newcommand{\tmu}{\tilde{\mu}}

\reversemarginpar

\begin{document}

{\par \texttt{SLAC-PUB-15430}\par}
{\par \texttt{CP3-Origins-2013-9 DNRF90}\par}
{\par \texttt{DIAS-2013-9}\par}
\bigskip{}

\title{Systematic Scale-Setting to All Orders:
\\
The Principle of Maximum Conformality and Commensurate Scale Relations 
}

\author{Stanley J. Brodsky}
\email{sjbth@slac.stanford.edu}
\affiliation{SLAC National Accelerator Laboratory, Stanford University, Stanford, California 94039, USA}

\author{Matin Mojaza}
\email{mojaza@cp3-origins.net}
\affiliation{CP3-Origins, Danish Institute for Advanced Studies, University of Southern Denmark, DK-5230}
\affiliation{SLAC National Accelerator Laboratory, Stanford University, Stanford, California 94039, USA}

\author{Xing-Gang Wu}
\email{wuxg@cqu.edu.cn}
\affiliation{Department of Physics, Chongqing University, Chongqing 401331, P.R. China}

\date{\today}

\begin{abstract}
We present in detail a new 
systematic method which can be used to automatically eliminate the
renormalization scheme and scale ambiguities in perturbative QCD predictions at all
orders.
We show that all of  the nonconformal $\beta$-dependent  terms in a QCD perturbative series can be readily identified by generalizing the conventional renormalization schemes based on dimensional regularization.  We then demonstrate that the nonconformal series of pQCD at any order can be resummed systematically into the scale of the QCD coupling in a unique and unambiguous way due to a special degeneracy of the $\beta$ terms in the series. 
The resummation follows from the principal of maximum conformality (PMC) and assigns
a unique scale for the running coupling at each perturbative order. 
The final result is independent of the initial choices of renormalization scheme and
scale, in accordance with the principles of the renormalization group, and thus eliminates an unnecessary source of systematic error in physical predictions.
We exhibit several examples known to order $\alpha_s^4$; i.e. 
i) the electron-positron annihilation into hadrons,
ii) the tau-lepton decay to hadrons,
iii) the Bjorken and  Gross-Llewellyn Smith (GLS) sum rules,
and iv) the static quark potential.
We show that the final series of the first three cases are all given in terms of
the anomalous dimension of the photon field in $SU(N)$, in accordance with conformality,
and with all non-conformal properties encoded in the running coupling.
The final expressions for the Bjorken and GLS sum rules directly
lead to the generalized Crewther relations, exposing another relevant feature of
conformality.
The static quark potential shows
that PMC scale setting in the Abelian limit is to all orders consistent with QED scale setting.
Finally, we demonstrate that the method applies to any renormalization scheme and can be used to derive commensurate scale relations between measurable effective charges,
which provide non-trivial tests of QCD to high precision.
This work extends BLM scale setting to any perturbative order,
with no ambiguities in identifying $\beta$-terms in pQCD, demonstrating that BLM scale setting follows from a principle of maximum conformality.
\begin{description}

\item[PACS numbers] 12.38.Aw, 12.38.Bx, 11.10.Gh, 11.15.Bt

\end{description}

\end{abstract}
\maketitle
\tableofcontents

\clearpage
\section{Introduction}
An important goal in high energy physics is to make perturbative QCD (pQCD) predictions as precise as possible, not only to test QCD itself, but also to expose new physics beyond the standard model. 
Recently, we showed a systematic method to determine the argument of the running coupling order by order in pQCD, and in a way that can be readily automatized \cite{Mojaza:2012mf,Wu:2013ei}. 
The new method satisfies all of the principles of the renormalization group \cite{Brodsky:2012ms}, and it eliminates an unnecessary source of systematic error.
The resulting predictions for physical processes are independent of theoretical conventions such as the choice of renormalization scheme and the initial choice of renormalization scale. The resulting scales also  determine the effective number of quark flavors at each order of perturbation theory. The method can be applied to processes with multiple physical scales and is consistent with QED scale setting. 

In this paper we review the method in detail and provide the complete generalization.
We show several examples, based on observables recently 
published in the literature to four-loop order in perturbation theory.
Finally, we demonstrate that the method applies to any renormalization scheme and can be used to derive commensurate scale relations between
measurable effective charges \cite{Grunberg:1980ja,Grunberg:1982fw,Dhar:1983py,Brodsky:1994eh}.
The method extends the Brodsky-Lepage-Mackenzie (BLM) method \cite{Brodsky:1982gc} to any perturbative order by following the Principle of Maximum Conformality
\cite{Brodsky:2011ig, Brodsky:2011ta,Brodsky:2012rj},
without leaving any ambiguity in identifying $\beta$-terms at any order in pQCD.

Previous attempts made in the literature to 
extend BLM scale setting to higher orders
\cite{Lu:1991yu,Lu:1991qr,Grunberg:1991ac,Grunberg:1992mp,Beneke:1994qe,Neubert:1994vb,LovettTurner:1995ti,Ball:1995ni,Brodsky:1997vq,Brodsky:2000cr,Hornbostel:2002af},
have mostly focused on improving convergence of the perturbative series and not removing renormalization scheme and scale ambiguities.
Therefore they do not in general satisfy the self-consistency requirements of the renormalization group \cite{Brodsky:2012ms}, nor the initial renormalization scale dependence, which must be the prerequisite of any scale-setting method. 
The main contribution of this work is to provide the systematic
method to eliminate the renormalization scale and scheme ambiguities to all orders in pQCD.

Other recent proposals similar in spirit are suggested in Refs.~\cite{Mikhailov:2004iq,Kataev:2010du}.

We start our analysis in Sec.~\ref{RdeltaScheme} by introducing a generalization of the conventional  schemes used in dimensional regularization in which a constant $-\delta$ is subtracted in addition to the standard  subtraction  \mbox{$\ln 4 \pi - \gamma_E$} of the $\MSB$-scheme. 
The $\delta$-subtraction defines an infinite set of renormalization schemes which we call \mbox{$\delta$-$\cal R$enormalization ($\R_\delta$)} schemes; since physical results cannot depend on the choice of scheme, predictions must be independent of $\delta$. 
As will be described in Sec.~\ref{resummation}, the $\R_\delta$-scheme exposes the general pattern of nonconformal $\{\beta_i\}$-terms, and it reveals a special degeneracy of the terms in the perturbative coefficients which allows us to resum the perturbative series. The resummed series matches the conformal series, which is itself free of any scheme and scale ambiguities
as well as being free of a divergent ÒrenormalonÓ series. It is the final expression one should use for physical predictions. It also makes it possible to set up an algorithm for automatically computing the conformal series and setting the effective scales for the coupling at each perturbative order.

In Sec.~\ref{Examples} we provide several examples, based on observables recently 
published in the literature to order $\alpha_s^4$; i.e. 
i) the electron-positron annihilation into hadrons,
ii) the tau-lepton decay to hadrons,
iii) the Bjorken and  Gross-Llewellyn Smith (GLS) sum rules,
and iv) the static quark potential.
We show explicitly that the final series of the first three cases are all given in terms of
the anomalous dimension of the photon field in $SU(N)$, in accordance with conformality,
and with all non-conformal properties encoded in the running coupling.
Moreover, the final expressions for the Bjorken and GLS sum rules directly
lead to the generalized Crewther relations, exposing another relevant feature of
conformality.
The static quark potential furthermore provides an example of how the method can be
automatized to give the PMC prediction directly from the number of quark flavor 
dependence of the initial expression. From this example we also demonstrate that the PMC prediction in the Abelian limit is consistent with QED scale setting.
Finally, we demonstrate in Sec.~\ref{CSR} that the method applies to any renormalization scheme and can be used to derive commensurate scale relations between measurable effective charges,
which provide non-trivial tests of QCD to high precision.

\section{The $\delta$-$\cal R$enormalization Scheme}\label{RdeltaScheme}
In dimensional regularization logarithmically divergent integrals are regularized by computing them in $d=4-2\epsilon$ dimensions \cite{Bollini:1972ui,'tHooft:1972fi,Bollini:1972bi}. 
This requires the following transformation of the integration measure
and introduction of an arbitrary mass scale $\mu$:
\ea{
\label{integral}
\int \rd^4 p \to  \mu^{2\epsilon} \int \rd^{4-2\epsilon}p  \ .
}
Divergences are then separated as $1/\epsilon$ poles and can be absorbed into redefinitions of the couplings. The choice of subtraction procedure is known as the \emph{renormalization scheme} and is chosen at the theorist's convenience. 
To avoid dealing with coupling constants changing dimensionality as a function of 
$\epsilon$ one rescales the couplings as well with the 
mass scale $\mu$ in the $d=4-2\epsilon$ theory.
In particular, for QCD one rewrites the bare gauge coupling $a_0 = \alpha_{0}/4\pi = g^2/(4\pi)^2$ as:
\begin{align}
\label{a0}
a_0 = \mu^{2\epsilon} Z_{a_\s} a_\s ,
\end{align}
where $a_\s$ is the \emph{renormalized} gauge coupling under a specific renormalization scheme $\s$ and $Z_{a_\s}$ is the renormalization constant of the coupling. The mass scale $\mu$ is now understood as the \emph{renormalization scale}.
The bare coupling must be independent of the arbitrary scale $\mu$, thus
\begin{align}
\mu^2 \frac{d a_0}{d\mu^2} = 0 .
\end{align}
Using this and the expansions
\begin{align}
\label{betadef}
\mu^2\frac{d a_\s}{d\mu^2} &= -\epsilon a_\s + \beta(a_\s)  \ , 
\\ 
\label{beta}
 \beta(a) &= - a^2 \sum_{i=0}^\infty \beta_i a^i \ ,
\\ 
 Z_a &= 1 + \sum_{i=1}^\infty z_i a^i \ , 
\end{align} 
it is easily derived that:
\begin{align}
\label{Z}
Z_a = &1 -\frac{\beta_0}{\epsilon} a + \left(\frac{\beta_0^2}{\epsilon^2}-\frac{\beta_1}{2\epsilon} \right) a^2 
\\
& - \left(\frac{\beta_0^3}{\epsilon^3} -\frac{7}{6}\frac{\beta_0\beta_1}{\epsilon^2} + \frac{\beta_2}{3\epsilon} \right )a^3 
\nn
&
+\left( \frac{\beta _0^4}{\epsilon ^4}-\frac{23 \beta _1 \beta _0^2}{12 \epsilon ^3}+\frac{5 \beta
   _2 \beta _0}{6 \epsilon ^2}+\frac{3 \beta _1^2}{8 \epsilon ^2}-\frac{\beta _3}{4
   \epsilon }\right) a^4 + \cdots  \ ,
   \nonumber
\end{align}
and the $\beta_i$ coefficients are known up to $\beta_3$, or four loops \cite{vanRitbergen:1997va}. The coefficients $\beta_i$ are renormalization-scheme dependent; however, it is easy to demonstrate by a general scheme-transformation that the first two coefficients $\beta_0$ and $\beta_1$ are universal for all mass-independent renormalization schemes.

In the minimal subtraction (MS) scheme \cite{'tHooft:1973mm} one absorbs the $1 / \epsilon$ poles appearing in loop integrals which come in powers of
\ea{
\ln \frac{\mu^2}{\Lambda^2} + \frac{1}{\epsilon} + c \ ,
}
where $c$ is the finite part of the integral.
Since anything can be hidden into infinity, one can subtract any finite part as well with the pole.
This is equivalent to redefining the arbitrary scale $\mu$ in Eq.~\eqref{integral}.
The $\overline{\rm MS}$-scheme \cite{Bardeen:1978yd} differs from the MS-scheme by an additional absorption of the term $\ln(4 \pi) - \gamma_E$,
which corresponds to redefining $\mu$ to:
\ea{
\mu^2 = \mu_{\MSB}^2 \ \exp({\ln 4 \pi - \gamma_E} )\ .
}

We will generalize this by defining the \mbox{ \it $\delta$-$\cal R$enormalization scheme}, $\cal R_\delta$, where one absorbs $\ln(4 \pi) - \gamma_E-\delta$, i.e.
\ea{
\mu^2 = \mu_\delta^2 \ \exp({\ln 4 \pi - \gamma_E-\delta}) \ ,
}
where $\delta$ is an arbitrary finite number, and by appropriate choice will connect all MS-type schemes.
In particular%
\footnote{Note that we have chosen $\MSB$ as the reference scheme for $\R_0$.
This is done since most results today are known in this scheme; however there is nothing special about $\MSB$, and $\R_0$ can be redefined to be any other MS-like scheme}:
\sea{
&{\cal R}_0 = \MSB \ , \\
&{\cal R}_{\ln 4\pi-\gamma_E} = {\rm MS} \ , \\
&{\cal R}_{-2} = {\rm G} \ ,
}
where we also provided the connection to the $G$-scheme, which
is yet another MS-like scheme proposed in the literature~\cite{Chetyrkin:1980pr}.

The scheme-transformation between different $\cal R_\delta$ corresponds simply to a displacement in their corresponding scales, 
i.e.
\begin{align}
\mu_{\delta_2}^2 = \mu_{\delta_1}^2 \exp({\delta_2 - \delta_1}) \ .
\end{align}
In particular:
\begin{align}
\label{scalerelations}
\mu_\delta^2 &= \mu_{\overline{\rm MS}}^2\  \exp({\delta}) \ . 
\end{align}

Since all $\cal R_\delta$'s are connected by scale-displacements, the $\beta$-functions of $a_{\R_\delta}$ defined in Eq. \eqref{betadef} are the same for all $\cal R_\delta$ to any order.
The index $\delta$ on $a_{\R_\delta}$ is thus redundant and we denote it instead as $a_\R$. Where it will not be ambiguous, we will simply use $ a \equiv a_\R$.

We can find a power series solution in $1/\ln(\mu/\Lambda)$ for $a$ by solving the renormalization group equation perturbatively. It is simplest to use the extended renormalization group prescription \cite{Stevenson:1981vj,Lu:1992nt} where one works with
a rescaled coupling and a rescaled
logarithm; respectively 
$$\hat{a} = \frac{\beta_1}{\beta_0} a \ , \quad  L_\delta = \frac{\beta_0^2}{\beta_1}\ln(\mu_\delta/\Lambda) \ . $$ 
The solution up to ${\cal O} (L_\delta^{-5})$ reads \cite{Brodsky:2011ta,Wu:2013ei}:
\begin{widetext}
\begin{eqnarray}\label{alphas}
\hat{a}(\mu_\delta) &=& \frac{1}{L_\delta}+ \frac{1}{L_\delta^2}\left({\cal C}- \ln L_\delta\right) + \frac{1}{L_\delta^3}\left[{\cal C}^2 +{\cal C} +c_2 -(2{\cal C}-\ln L_\delta +1)\ln L_\delta -1\right] + \frac{1}{L_\delta^4}\left\{ {\cal C}\left({\cal C}^2 +\frac{5}{2}{\cal C} + 3 c_2 -2\right) \right.\nonumber\\
&& \left. -\frac{1-c_3}{2} -\left[3{\cal C}^2 +5{\cal C} +3c_2-2 -\left(3{\cal C} -\ln L_\delta +\frac{5}{2}\right)\ln L_\delta \right]\ln L_\delta\right\} +{\cal O}\left(\frac{1}{L_\delta^5}\right) ,
\end{eqnarray}
\end{widetext}
where $c_i = \beta_i \beta_0^{i-1}/\beta_1^i$ are the rescaled $\beta$-function coefficients and $\cal C$ is an arbitrary integration constant which in $\R_\delta$ is set to
${\cal C } = \ln \beta_0^2/\beta_1$ in order to reproduce the standard $\Lambda_\MSB$ scale \cite{Lu:1992nt, Brodsky:2011ta,Wu:2013ei} . Note that we take the asymptotic scale $\Lambda=\Lambda_\MSB$ to be the same for any $\R_\delta$. Alternatively, one can take the scale $\mu$ to be the same for any $\R_\delta$, having instead different asymptotic scales $\Lambda_\delta$.

This solution for $\hat{a}(\mu_\delta)$ holds for massless QCD or
as long as the active quark masses are below $\Lambda$.
In the case where there are active quarks with masses higher than $\Lambda$,
one must take the quark threshold effects into account when running
the coupling from $\Lambda$ to $\mu_\delta$.
This can be done by e.g. solving the renormalization group equation
with an analytic $\beta$-function that takes quark masses into account \cite{Brodsky:1998mf} or by using matching equations at each 
quark mass threshold when running the coupling to $\mu_\delta$ \cite{Chetyrkin:1997sg}.

\section{Observables in $\R_\delta$}\label{resummation}
Consider an observable in pQCD in some scheme which we put as the reference scheme $\R_0$ with the following expansion:
\ea{
\label{observable}
\rho_0 (Q^2)= a(\mu_0)^n \sum_{k=0}^\infty r_{k+1} (Q^2/\mu_0^2) a(\mu_0) ^k \ ,
}
{where $\mu_0$ stands for some initial renormalization scale and $Q$ is the kinematic scale of the process}. 
The full pQCD series is formally 
 independent of the choice of the initial renormalization scale $\mu_0$, 
 if it were possible to sum the entire series.
 However, this goal is not feasible in practice, especially because of the $k! \beta^k \alpha_s^k$ renormalon growth of the nonconformal series. When a perturbative expansion is truncated at any finite order, it generally becomes renormalization-scale and scheme dependent; i.e.,  dependent on theoretical conventions.
This can be exposed by using the $\R_\delta$-scheme.
Since results in any $\R_\delta$ are related by scale displacements,
we can derive the general expression for $\rho$ in $\R_\delta$
by using the displacement relation between couplings in any $\R_\delta$-scheme:
\ea{
\label{adelta}
a(\mu_0) = a(\mu_\delta) + \sum_{n=1}^\infty \frac{1}{n!} { \frac{{\rm d}^n a(\mu)}{({\rm d} \ln \mu_0^2)^n}\left|_{\mu=\mu_\delta}\right. (-\delta)^n} \ ,
}
{where we used $\ln\mu^2_0/\mu^2_\delta=-\delta$}. 
It is useful to derive the general displacement relation for $a(\mu_0)^k$ for any $k$ as an expansion in $a$ up to order $a^{k+3}$:
\ea{
&
a(\mu_0)^k = a(\mu_\delta)^k + k \beta_0 \delta a(\mu_\delta)^{k+1} 
\\ &
+ k\left[ \beta_1 \delta + \frac{k+1}{2} \beta_0^2 \delta^2 \right] a(\mu_\delta)^{k+2}
\nn 
&
+k\left [\beta_2 \delta  + \frac{2k+3}{2} \beta_0\beta_1\delta^2 + \frac{(k+1)(k+2)}{3!} \beta_0^3 \delta^3 \right] a(\mu_\delta)^{k+3} .
\nonumber
}
Inserting this expression into Eq.~\eqref{observable} at each order $a(\mu_0)^k$
we find the expression for $\rho$ for an arbitrary $\delta$ to order $a^4$, that is in any $\R_\delta$-scheme, to be:
\begin{widetext}
\begin{align}
\label{rhodeltas}
\rho_\delta (Q^2) = & r_1 a_1(\mu_\delta)^n +
[r_2 + n \beta_0 r_1 \delta_1] a_2(\mu_\delta)^{n+1}
+\left [r_3 +n \beta _1 r_1\delta_1 +(n+1)\beta _0 r_2 \delta_2+ \frac{n(n+1)}{2} \beta _0^2 r_1 \delta_1^2\right ] a_3(\mu_\delta)^{n+2}
\nn
&
+\left [ r_4 +n \beta _2 r_1\delta_1
 +(n+1)\beta _1 r_2 \delta_2 + (n+2)\beta_0 r_3 \delta_3 
+ \frac{n(3+2n)}{2}Ê\beta_0\beta_1 r_1 \delta_1^2  + \frac{(n+1)(n+2)}{2} \beta_0^2 r_2 \delta_2^2
 \right .
\nn & 
\hspace{9.5mm} \left . 
+ \frac{n(n+1)(n+2)}{3!} \beta_0^3 r_1 \delta_1^2 \right ] a_4(\mu_\delta)^{n+3} 
+ {\cal O}(a^5) \ ,
\end{align}
\end{widetext}
where $\mu_\delta^2 = Q^2 e^{\delta}$, the initial scale is for simplicity set to $\mu_0^2 = Q^2$, and we have defined in Eq.~\eqref{observable} $r_i(1) = r_i$. An artificial index was introduced on each $a$ and $\delta$ to keep track of which coupling each $\delta$ term is associated with. 
They are not an indication of different variables; i.e. $a_1 = a_2 = \ldots$ and $\delta_1 = \delta_2 = \ldots $. 
The use of the artificial indices will be made clear in a moment.

The above expression shows the scheme dependence explicitly;
e.g. if \mbox{$\R_0 = \MSB$}, then choosing 
\mbox{$\delta = \delta_i = \ln 4 \pi- \gamma_E $}
will give the result in the MS-scheme. 
The initial scale choice is arbitrary and is not the final argument of the running coupling; the final  scales will be independent of the initial renormalization scale.

In a  conformal (or scale-invariant) theory, where $\{\beta_i\}=\{0\}$, the $\delta$ dependence vanishes in Eq.~\eqref{rhodeltas}. 
Therefore by absorbing all $\{\beta_i\}$ dependence into the running coupling at each order, we obtain a final result independent of the initial choice of scale and scheme. The coefficients in the final expression will thus be equal 
to those of the conformal theory.
The use of $\R_\delta$ allows us to put this on rigorous grounds.
From the explicit expression in Eq.~\eqref{rhodeltas} it is easy to confirm that
\begin{align}
\frac{\partial \rho_\delta}{\partial \delta} = -\beta(a) \frac{\partial \rho_\delta}{\partial a} \ .
\end{align}
The scheme-invariance of the physical prediction requires that $\partial \rho_\delta /\partial \delta = 0$. 
Therefore the scales in the running coupling must be shifted and set such that the conformal terms associated with the $\beta$-function are removed;  the remaining conformal terms are by definition renormalization scheme independent.
The numerical value for the prediction at finite order is then scheme independent as required by the renormalization group.
The scheme-invariance criterion is a theoretical requirement of the renormalization group; it must be satisfied at any truncated order of the perturbative series, and is different from the formal statement that the all-orders expression for a physical observable is renormalization scale and scheme invariant; i.e. ${\rm d} \rho /{\rm d} \mu_0 = 0$.
The final series obtained corresponds to the theory for which
\mbox{$\beta(a) = 0$}; i.e. the conformal series.
This demonstrates to any order the concept of the \emph{principal of maximum conformality} (PMC) \cite{Brodsky:2011ta}, which states that all non-conformal terms in the perturbative series must be resummed into the running coupling.

The expression in Eq.~\eqref{rhodeltas} exposes the pattern of $\{\beta_i\}$-terms in the coefficients at each order.
It is possible to infer more from Eq.~\eqref{rhodeltas}; since
 there is nothing special about a particular value of $\delta$,
we conclude that some of the coefficients of the $\{\beta_i\}$-terms are degenerate; 
e.g. the coefficient of $\beta_0 a(Q)^2$ and $\beta_1 a(Q)^3$ can be set equal.
Thus for any scheme, the expression for $\rho$ can be put to the form:
\begin{widetext}
\ea{
\rho(Q^2) = &r_{1,0} a(Q)^n + [r_{2,0} + n \beta_0 r_{2,1} ] a(Q)^{n+1}   + \left  [r_{3,0} +
 n\beta_1 r_{2,1} + (n+1) \beta_0 r_{3,1} + \tiny \frac{n(n+1)}{2} \beta _0^2 r_{3,2} \right ]a(Q)^{n+2}  \nn
&
+\left [ r_{4,0} +n \beta _2 r_{2,1} +(n+1)\beta _1 r_{3,1} + (n+2)\beta_0 r_{4,1} 
+ \frac{n(3+2n)}{2}Ê\beta_0\beta_1 r_{3,2}  + \frac{(n+1)(n+2)}{2} \beta_0^2 r_{4,2}
\right .
\nn
& 
\left . \qquad
+ \frac{n(n+1)(n+2)}{3!} \beta_0^3 r_{4,3} \right ] a(Q)^{n+3} + {\cal O}(a^{n+4})\ ,
\label{betapattern}
}
where the $r_{i,0}$ are the conformal parts of the perturbative coefficients; i.e. 
$r_{i} = r_{i,0} + {\cal O}(\{\beta_i\})$. 
\end{widetext}
The $\R_\delta$-scheme not only illuminates the $\{\beta_i\}$-pattern, but it also exposes a \emph{special degeneracy} of coefficients at different orders.
The degenerate coefficients can themselves be functions of $\{\beta_i\}$, hence
Eq.~\eqref{betapattern} is not to be understood as an expansion in $\{\beta_i\}$, but
at pattern of $\{\beta_i\}$ with degenerate coefficients that must be matched.

The artificial indices in the expansion in Eq.~\eqref{rhodeltas} reveals how the $\{\beta_i\}$-terms must be absorbed into the running coupling: The different $\delta_k$'s keep track of the 
power of the $1/\epsilon$ divergence of the associated diagram at each loop order in the following way; the $\delta_k^p a^{m}$-term indicates the term associated to 
a diagram with $1/\epsilon^{1+m-n-k}$ divergence for any power $p$ of $\delta$.
Grouping the different $\delta_k$-terms, one recovers
in the $N_c \to 0$ Abelian limit \cite{Brodsky:1997jk} the dressed skeleton expansion.
Resumming the series according to this expansion thus
correctly reproduces the QED limit of the observable
and matches the conformal series with the running coupling
evaluated at effective scales at each order.
Using this information we can rearrange
the expression in Eq.~\eqref{betapattern} in 
the skeleton-like expansion:
\begin{widetext}
\ea{
\rho(Q^2) = &a(Q)^n \Big[r_{1,0}   + n \left(\beta_0 a(Q) + \beta_1 a(Q)^2 + \beta_2 a(Q)^3 \right) r_{2,1} 
+ \frac{n}{2} \left ((n+1) \beta _0^2 a(Q)^2
 + (3+2n)Ê\beta_0\beta_1 a(Q)^3 \right ) r_{3,2}
 \nn
 &
 \qquad \quad+ \frac{n(n+1)(n+2)}{3!} \beta_0^3 r_{4,3} a(Q)^3 \Big] 
\nn
&
 +   
a(Q)^{n+1} \left [r_{2,0} + (n+1) \left(\beta_0 a(Q) + \beta_1 a(Q)^2 \right) r_{3,1} 
+ \frac{(n+1)(n+2)}{2} \beta_0^2 r_{4,2} a(Q)^2 \right ]
\nn
&
+
a(Q)^{n+2} \big [r_{3,0} + (n+2)\beta_0 r_{4,1} a(Q) \big]
\nn
&
+
a(Q)^{n+3}\big [ r_{4,0} \big ]   + {\cal O}(a^{n+4})\ . 
\label{skeletonexpansion}
}
\end{widetext}


\subsection{Systematic All-Orders PMC scale setting}\label{Systematic}
It is easy to see from Eq.~\eqref{skeletonexpansion} that we can resum all $r_{i, 1}$ terms, which come with a linear factor of $\beta_j$, \emph{to all orders} by defining new scales $Q_i$ at each order as follows (for simplicity, we treat the higher-power $\beta_j$ terms later):
\begin{align}
r_{1,0} a(Q_1)^n &= r_{1,0} a(Q)^n - n a(Q)^{n-1}\beta(a) r_{2,1}  \ ,
\nn
r_{2,0} a(Q_2)^{n+1} &= r_{2,0} a(Q)^{n+1} - (n+1) a(Q)^n \beta(a) r_{3,1} \ ,
\nn
r_{3,0} a(Q_3)^{n+2} &= r_{3,0} a(Q)^{n+2} - (n+2) a(Q)^{n+1} \beta(a) r_{4,1} \ ,
\nn
&\hspace{2mm} \vdots 
\nn
r_{k,0} a(Q_k)^{k} &= r_{k,0} a(Q)^k - k \ a(Q)^{k-1} \beta(a) r_{k+1,1} .\label{PMCscale1}
\end{align}
From the scale displacement equation \eqref{adelta} for $a$ it
is straightforward to see that:
\ea{
\label{aQk}
&a(Q_k)^k = a(Q)^k + k a(Q)^{k-1} \beta(a) \ln \frac{Q_k^2}{Q^2} +
\\
&+ \frac{k}{2} a(Q)^{k-2}\left [  \beta \frac{\partial \beta}{\partial a} a(Q) + (k-1) \beta(a)^2 \right ] \ln^2 \frac{Q_k^2}{Q^2} + \cdots \ .
\nonumber
}
It follows from identifying Eq.~\eqref{PMCscale1} with \eqref{aQk} that to absorb all linear $\beta_j$ terms, the
scales $Q_k$ must satisfy:
\ea{ - \frac{r_{k+1,1}}{r_{k,0}} = \ln \frac{Q_k^2}{Q^2} +\frac{1}{2}\left [ \frac{\partial\beta}{\partial a} + (k-1) \frac{\beta}{a} \right] \ln^2 \frac{Q_k^2}{Q^2} + \cdots \ , 
}
where $r_{k,0}$ are the conformal coefficient and $r_{k+1,1}$ are the degenerate coefficients of linear $\beta_j$-terms.
This leads to the self-consistency equation for $Q_k$:
\ea{
\ln \frac{Q_k^2}{Q^2}  = \frac{- r_{k+1,1}/r_{k,0}}{1+ \frac{1}{2}\left [ \frac{\partial\beta}{\partial a} + (k-1) \frac{\beta}{a} \right] \ln\frac{Q_k^2}{Q^2}  + \cdots } \ .
}
\\[-2mm]

\noindent
To leading logarithmic order (LLO) we have:
\ea{
\ln \frac{Q_{k,LLO}^2}{Q^2}  = -\frac{r_{k+1,1}}{r_{k,0}} \ .
}
This resums all linear $\beta_j$ terms, but introduces higher-power $\beta_j$ terms beyond the order $a^{k+1}$. 
For example, suppose that we are computing an observable to order $a^p$. 
The scales $Q_k$ must resum all $\beta_j r_{k+1,1}$ terms without introducing higher order ones up to order $a^p$. This means that $Q_k$ must be computed to the ${p-(k+1)}$ logarithmic order (N${}^{p-(k+1)}$LLO).
Let us explicitly perform the resummation up to $a^4$ for the first scale $Q_1$, that is, up to next-to-next-to-leading logarithmic order (NNLLO).
The general expression for the NLLO scale reads:
\ea{
\label{QkNLLO}
\ln \frac{Q_{k,NLLO}^2}{Q^2}  = \frac{- r_{k+1,1}/r_{k,0}}{1+ \frac{1}{2}\left [\frac{\partial\beta}{\partial a} + (k-1) \frac{\beta}{a} \right]\left ( -\frac{r_{k+1,1}}{r_{k,0}}\right)} \ .
}
To find the NNLLO scale, we first write the self-consistency equation (exposing one higher logarithmic order in the denominator):
\begin{widetext}
\ea{
\label{SCNNLO}
\ln \frac{Q_k^2}{Q^2}  = \frac{- r_{k+1,1}/r_{k,0}}{1+ \frac{1}{2}\left [ \frac{\partial\beta}{\partial a} + (k-1) \frac{\beta}{a} \right] \ln\frac{Q_k^2}{Q^2}  
+ \frac{1}{3!}\left [   \beta \frac{\partial^2 \beta}{\partial a^2} + \left( \frac{\partial\beta}{\partial a} \right )^2 + 3(k-1) \frac{\beta}{a}\frac{\partial\beta}{\partial a} + (k-1)(k-2) \frac{\beta^2}{a^2}\right ] \ln^2\frac{Q_k^2}{Q^2} + \cdots} \ .
}
Then we replace the
logarithms in the denominator with the expansion of its NLLO expression in Eq.~\eqref{QkNLLO}:
\ea{
\ln \frac{Q_{k,NLLO}^2}{Q^2} = -\frac{r_{k+1,1}}{r_{k,0}}\left ( 1 + \frac{1}{2}\left [ \frac{\partial\beta}{\partial a} + (k-1) \frac{\beta}{a} \right]\frac{r_{k+1,1}}{r_{k,0}}  + \cdots \right) \ .
}
We thus get:
\ea{
\ln \frac{Q_{k,NNLLO}^2}{Q^2}  = \frac{- r_{k+1,1}/r_{k,0}}{1+ \frac{1}{2}\left [ \frac{\partial\beta}{\partial a} + (k-1) \frac{\beta}{a} \right] \left ( -\frac{r_{k+1,1}}{r_{k,0}}\right) 
+ \frac{1}{3!}\left [  \beta \frac{\partial^2 \beta}{\partial a^2} - \frac{1}{2} \left( \frac{\partial\beta}{\partial a} \right )^2 - \frac{(k-1)(k+1)}{2} \frac{\beta^2}{a^2}\right ] \left ( \frac{r_{k+1,1}}{r_{k,0}}\right)^2} \ .
}
This procedure iterates to any desired order.
\end{widetext}

For observables, where the higher-power $\beta_j$ coefficients vanish,
i.e. $r_{3,2}=r_{4,2}=r_{4,3}=0$ (this is the case in e.g. 
the Adler $D$-function), these scales give the final PMC expression for the observable, which is invariant under any
scheme transformation:
\ea{
\rho(Q^2) = & r_{1,0} a(Q_{n,NNLLO})^n + r_{2,0} a(Q_{n+1,NLLO})^{n+1}
\nn
&+r_{3,0} a(Q_{n+2,LLO})^{n+2} + r_{4,0}a(Q)^{n+3}+{ \cal O }(a^{n+4}) \ .
}
This is the conformal series with coefficients that are independent of the renormalization scheme.
Note that the last scale remains ambiguous.
This ambiguity only affects the highest order term. The final expression and
coefficients are therefore not affected by the ambiguity of the last scale and thus the renormalization scheme dependence has been eliminated and the
renormalization scale dependence only resides in the highest power coupling of the perturbative series.
We note that one does not need the full expression of the $a^5$ coefficient to set the last scale, $Q_{n+3}$, but only the coefficient $r_{5,1}$. 

Let us now generalize to observables that do depend on higher powers in $\beta_j$. This is for example the case in $R_{e^+e^- \to \rm hadrons}(s)$.
One can use the procedure just describe, but instead
of Eq.~\eqref{PMCscale1} we use its generalization:

\begin{widetext}
\begin{align}
r_{k,0} a(Q_k)^{k} &= r_{k,0} a(Q)^k - k \ a(Q)^{k-1} \beta(a) r_{k+1,1} + 
\frac{k}{2} \left [ a(Q)^{k-1} \frac{\rd\beta}{\rd \ln\mu^2}+ (k-1)a(Q)^{k-2}\beta(a)^2 \right ] r_{k+2,2} + \cdots \ .
\label{PMCscale}
\end{align}
It is easy to verify that these expressions, which define the PMC scales $Q_k$, correctly resum all $\{\beta_i\}$-terms in $\rho$. 
Eq.~\eqref{PMCscale} is systematically derived by replacing the $\ln^j Q_1^2/Q^2$ by $r_{k,j}$ in the logarithmic expansion of $a(Q_k)^k$ in Eq.~\eqref{aQk} up to the highest known $r_{k,n}$-coefficient in pQCD.
We introduce a short-hand notation of Eq.~\eqref{PMCscale}:
\ea{
a(Q_k)^{k} &= a(Q)^k + k \ a(Q)^{k-1} \beta(a) \left \{ 
R_{k,1} +\Delta_k^{(1)}(a) R_{k,2} + \Delta_k^{(2)}(a) R_{k,3} + \cdots + \Delta_k^{(n)}(a) R_{k,n+1} \right \} \ ,
}
where 
\sea{
R_{k,j} &= (-1)^{j}\frac{r_{k+j, j}}{r_{k,0}} \ , \\
\Delta_k^{(1)}(a) &= \frac{1}{2} \left [ \frac{\partial \beta}{\partial a} + (k-1) \frac{\beta}{a}\right] \ , \\
\Delta_k^{(2)}(a) & = 
 \frac{1}{3!}\left [   \beta \frac{\partial^2 \beta}{\partial a^2} +Ê \left( \frac{\partial\beta}{\partial a} \right )^2  + 3(k-1) \frac{\beta}{a}\frac{\partial\beta}{\partial a} + (k-1)(k-2) \frac{\beta^2}{a^2}\right ] \ , \  \ldots \ .
}
Following the same procedure as before, one finds the final expressions for $Q_{k,LLO}$, $Q_{k,NLLO}$ and $Q_{k,NNLLO}$ to be:
\seal{effectivescales}{
\ln \frac{Q_{k,LLO}^2}{Q^2}  &= R_{k,1} \ ,\label{LOscale}\\
\ln \frac{Q_{k,NLLO}^2}{Q^2}  &= \frac{R_{k,1} + \Delta_k^{(1)}(a) R_{k,2}}{1+ \Delta_k^{(1)}(a) R_{k,1}} \ , \\
\ln \frac{Q_{k,NNLLO}^2}{Q^2}  &= \frac{R_{k,1} + \Delta_k^{(1)}(a) R_{k,2}+\Delta_k^{(2)}(a) R_{k,3}}{1+ \Delta_k^{(1)}(a) R_{k,1} + \left({\Delta_k^{(1)}(a)}\right)^2 (R_{k,2} -R_{k,1}^2)  + \Delta_k^{(2)}(a)R_{k,1}^2 } 
 \ , \ \ldots . \label{exactscales}
}
\end{widetext} 
These final expression are generic and can be used directly.
We thus have a procedure which systematically sets the
PMC scales to all-orders.

It is easy to see that the leading order values
of the effective scales are independent of the initial renormalization scale $\mu_0$.
This follows since taking $\mu_0 \neq Q$ we must replace $R_{k,1} \to R_{k,1} + \ln Q^2/\mu_0^2$ and thus the leading order effective scales read
$\ln Q_{k, \rm LO}^2/\mu_0^2 = R_{k,1} + \ln Q^2/\mu_0^2$, where $\mu_0$ cancels and Eq.~\eqref{LOscale} at LLO is recovered. This generalizes to any order. 
Since the $\beta$-function is not known to all orders, 
 a higher order residual renormalization-scale dependence will enter through the running coupling. This residual renormalization-scale dependence is strongly suppressed in the perturbative regime of the coupling \cite{Brodsky:2012sz,Brodsky:2012ik}.

The effective scales contain all the information of the non-conformal parts of the initial pQCD expression for $\rho$ in Eq.~\eqref{betapattern}; 
this is exactly the purpose of the running coupling.
The quotient form of Eq.~\eqref{exactscales} sums up an infinite
set of terms related to the known $r_{j,k \neq 0}$ which appear at every higher order due to the special degeneracy of Eq.~\eqref{betapattern}. The method systematically sums up all known non-conformal terms, in principle to all-orders, but is in practice truncated due to the limited knowledge of the $\beta$-function.

In earlier PMC scale setting \cite{Brodsky:2012rj,Brodsky:2011ig,Brodsky:2011ta}, 
and its predecessor, the Brodsky-Lepage-Mackenzie (BLM) method \cite{Brodsky:1982gc,Grunberg:1991ac, Brodsky:1994eh},
the PMC/BLM scales have been set by using a perturbative expansion in $a$ and only approximate conformal series have been obtained. Here, we have been able to obtain the conformal series 
as revealed in dimensional regularization schemes. The final scales in Eq.~\eqref{effectivescales} have naturally become functions of the coupling through the $\beta$-function, in principle, to all orders.

\subsection{Automation}\label{Automation}
In many cases the coefficients in a pQCD expression for an observable are computed numerically, and
the $\{\beta_i\}$ dependence is not known explicitly. It is, however, easy to extract the dependence on the number of quark flavors $N_f$, since $N_f$ enters analytically in any loop diagram computation.
To use the systematic method presented in this letter one puts the pQCD expression into the form of Eq.~\eqref{betapattern}. Due to the special degeneracy in the coefficient of the $\{\beta_i\}$-terms, the $N_f$ series 
can be matched to the $r_{j,k}$ coefficients in a unique way%
\footnote{
In principle, one must treat the $N_f$  terms unrelated to renormalization of the gauge coupling as part of the conformal coefficient; e.g.,
the $N_f$ terms coming from light-by-light scattering in QED and the $N_f$  terms unrelated to the renormalization of the tri-gluon and quartic-four-gluon vertices belongs to the conformal series.
}.
This allows one to automate the
scale setting process algorithmically.

The $n$-th order coefficient in pQCD has an expansion in $N_f$ which reads:
\ea{
r_{n} = c_{n,0} + c_{n,1} N_f + \cdots + c_{n,n-1} N_f^{n-1} \ .
}
By inspection of Eq.~\eqref{betapattern} it is seen that there are exactly as many unknown coefficients in the $\{\beta_i\}$-expansion at the order $a^n$ as the $N_f$ coefficients, $c_{n,j}$. This is realized due to the special degeneracy found in \eqref{betapattern}.
The $r_{i,j}$ coefficients in Eq.~\eqref{betapattern} can thus be expressed in terms of the $c_{n,j}$ coefficients. 
The highest power in $N_f$ at any order should 
always be associated with the same power in $\beta_0$.
The first $\beta_0$ appears at order $a^{n+1}$.
We derive the relations between $c_{n,j}$ and $r_{i,j}$
for a general gauge group, where we define
$C_A$ and $C_F$ as the quadratic Casimir coefficients of
the adjoint and quark representations and
$T$ as the generator trace normalization. For QCD
these coefficients read: $C_A = N_c$, $C_F = (N_c^2-1)/2N_c$ and $T = 1/2$.
Using that $\beta_0 = 11/3 C_A -4/3 T N_f $ we can find $r_{2,0}$ and $r_{2,1}$:
\ea{
r_2 = r_{2,0} + n  \beta_0 r_{2,1} =
\left [ r_{2,0} + r_{2,1} \frac{11 n C_A}{3} \right ]
- r_{2,1} \frac{4n}{3} T N_f \ ,\nonumber 
}
This leads to:
\ea{
r_{2,1} = - \frac{3 }{4  T} \frac{c_{2,1}}{n} \ , \quad
r_{2,0}  = c_{2,0} + \frac{11 C_A}{4T} c_{2,1}  \ .
}
At the next order we have:
\ea{
r_3 = r_{3,0} + n  \beta_1 r_{2,1}+ (n+1) \beta_0 r_{3,1} + \frac{n(n+1)}{2}\beta_0^2 r_{3,2}  \ , 
\nonumber
}
where $r_{2,1}$ is already known.
Expanding as before in terms of $N_f$ (with the higher order $\beta_i$ coefficient given in \cite{vanRitbergen:1997va}) we find the matching:
\seal{r3coef}{
r_{3,2} &= \frac{9 }{8 T^2} \frac{c_{3,2}}{n(n+1)} \ , \\
r_{3,1}&= \frac{1}{8 (n+1) T^2}\left [6 T c_{2,1} \left(5 C_A+3 C_F\right)
  \right . \nn
   & \qquad \qquad \qquad \left .
-33 c_{3,2} C_A-6 T
   c_{3,1}\right ] \ , \\
  r_{3,0}&= c_{3,0}+ \frac{1}{16 T^2} \left [11 C_A \left(11
   c_{3,2} C_A+4 T c_{3,1}\right) 
   \right . \nn
   & \qquad \qquad \qquad \left .-12 T c_{2,1} C_A \left(7 C_A+11 C_F\right) \right ] \ .
   }

Similarly, we can find the $r_{4,j}$ coefficients:
  \begin{widetext}
\seal{r4coef}{
r_{4,3} &= \left(- \frac{3 }{4 T} \right )^3 \frac{3!}{n(n+1)(n+2)}c_{4,3}  \ , \\
r_{4,2} &=\frac{1}{32(n+1)(n+2) T^3} \left [2 n T^2 c_{2,1} \left(79 C_A+66 C_F\right)-9 \left(\frac{4 (3+2n)}{n+1} T c_{3,2}
   \left(5 C_A+3 C_F\right)-33 c_{4,3} C_A-4 T c_{4,2}\right) \right ]  \ , \\
r_{4,1} &= \frac{1}{64 (n+2) T^3} \Bigg [
4  T^2 c_{2,1} \left(-(37 n+360) C_A C_F+2 (91 n-150)
   C_A^2-18 (n+6) C_F^2\right)
   +48 T^2 c_{3,1} \left(5 C_A+3
   C_F\right)
       \nn
   & \qquad 
   +\frac{12 T c_{3,2}}{n+1} C_A \big((152 n+173) C_A+33 (4 n+5)
   C_F \big)
   - 33 C_A \left(33 c_{4,3} C_A+8 T c_{4,2}\right)-48
   T^2 c_{4,1}
 \Bigg]  \ , \\
r_{4,0}&= c_{4,0}+
\frac{1}{64T^3} \Big [
2 T^2 c_{2,1} C_A \left(8 (228-77 n) C_A C_F+(840-1127 n)
   C_A^2+132 (n+6) C_F^2\right)
    -48 T^2 c_{3,1} C_A \left(7 C_A+11 C_F\right)
       \nn
   & \qquad \qquad \qquad  
  -2904 T c_{3,2} C_A^2 C_F+176 T^2 c_{4,1} C_A-1848 T c_{3,2}
   C_A^3
   +484 T c_{4,2} C_A^2+1331 c_{4,3} C_A^3
 \Big ]  \ .
 }
Using these relations
automatically gives the effective scales
in Eq. \eqref{exactscales}.

 \end{widetext}

The automation process can be outlined as follows:
\begin{enumerate}
\item Choose any $\delta$-$\cal R$enormalization scheme and scale.
\item Compute the physical observable in pQCD and extract the $N_{\rm f}$ coefficients, $c_{k,j}$. 
\item Find the $\beta_i$ coefficients, $r_{k,j}$ from the $c_{k,j}$ coefficients and compute the PMC scales, $Q_k$.
\item The final pQCD expression for the observable reads 
\ea{
\rho_{\rm final} (Q) = \sum_{k=0} r_{k+1,0} a(Q_{k+1})^{n+k} \ .
}
\end{enumerate}

This procedure demonstrates that the $N_f$ terms can be unambiguously associated to the $\{\beta_i\}$-terms to all orders. It also shows that PMC is the underlying principle of BLM scale setting.

The PMC method can be used to set separate scales for different skeleton diagrams; this is particularly important for multi-scale processes. In general the $\{\beta_i\}$-coefficients multiply terms involving logarithms in each of the invariants \cite{Brodsky:2011ig}.
For instance, in the case of $q \bar q \to Q \bar Q$ near the heavy quark threshold in pQCD, the PMC assigns different scales to the annihilation process and the rescattering corrections involving the heavy quarks' relative velocity \cite{Brodsky:1995ds}.
It also can be used to set the scale for the ``lensing" gluon-exchange corrections that appear in the Sivers, Collins, and Boer-Mulders effects.
Moreover, for the cases when the process involves several energy regions; e.g. hard, soft, etc., one may adopt methods such as the non-relativistic QCD effective theory (NRQCD) \cite{Bodwin:1994jh} and the soft-collinear effective theory (SCET) \cite{Bauer:2000yr,Bauer:2001yt} to set the PMC scales; i.e., one first sets the PMC scales for the higher energy region, then integrate it out to form a lower energy effective theory and sets the PMC scales for this softer energy region, etc. In this way one obtains different effective PMC scales for each energy region, at which all the PMC properties also apply. 

As an important remark, one should keep in mind that the determination of the factorization scale is a separate issue from renormalization scale setting since it is present even in a conformal theory when $\beta=0$. 
Nevertheless, in the literature the factorization scale in hadronic processes is often set
to be equal to the renormalization scale. 
In principle, the factorization scale can be determined if one has knowledge of the nonperturbative light-front wavefunctions of the initial or final state hadrons. However, the PMC can also  be used to set the scale of the coupling that appears in the DGLAP or ERBL evolution equations, and
it is consistent with the usual factorization properties for hard-process cross sections in QCD.
It is therefore important to separate the renormalization and factorization scales in hadronic processes~\cite{Wu:2013ei}.

\section{Examples}\label{Examples}
We now consider three examples based on 
the Adler function \cite{Adler:1974gd}, $D$,
which can be measured indirectly
through the dispersion relation:
\ea{
\label{AdlerDispersion}
D(Q^2) = Q^2 \int_{4 m_\pi^2}^\infty \frac{R_{e^+e^-}(s)}{(s+Q^2)^2} {\rm d}s \ , 
}
where $R_{e^+e^-}$ is the ratio for electron-positron annihilation 
into hadrons.

The Adler function is particularly instructive to consider, 
since its conformal and non-conformal parts
can be separated by using RG arguments.
Explicitly, the Adler function can
be written in terms of the photon field anomalous dimension,
$\gamma$, and the vacuum polarization function, $\Pi$, 
as follows  \cite{Chetyrkin:1996ia,Baikov:2012zm}
\begin{align}
\label{Adler}
\bar{D}(Q^2) = \kappa^{-1} D(Q^2) =  \gamma(a) - \beta(a) \frac{d}{da} \Pi(Q^2,a) \ .
\end{align}
where $\beta(a)$ is the $\beta$-function of the running coupling and we have defined the \emph{normalized} Adler function
$\bar{D}$ where $\kappa = d_F \sum_f Q_f^2$ and $d_F$ is the dimension
of the quark color representation, which in QCD reads $d_F = N_c$.
We will work with this normalization throughout the related examples.
In perturbation theory we define
\ea{
\gamma (a) &= \kappa \sum_{n=0}^\infty \gamma_n a(Q)^n \ ,\\
\Pi(a) & = \kappa \sum_{n=0}^\infty \Pi_n a(Q)^n \ ,
}
which are now known to four-loop order \cite{Chetyrkin:1979bj,Dine:1979qh,Celmaster:1979xr,Celmaster:1980ji,Gorishnii:1990vf,Surguladze:1990tg,Baikov:2012zm,Baikov:2008jh,Baikov:2010je,Baikov:2012zn}.
The PMC procedure then follows by
absorbing all $\beta$-dependent terms, which following Sec.~\ref{Systematic}
becomes a trivial exercise
once the degenerate coefficients $r_{i,j}$ have been identified.

As a fourth example, we consider a case where
the explicit conformal and non-conformal parts are not known.
Here we make explicit use of the automation procedure 
to derive the special degeneracy as described in Sec.~\ref{Automation}.

\subsection{$e^+e^- \to$ {hadrons}}
The ratio for electron-positron annihilation into hadrons,
$R_{e^+e^-}$ can inversely to Eq.~\eqref{AdlerDispersion}
be computed from the Adler function, $D$, as follows:
\begin{align}
\bar{R}_{e^+e^-}(s) = \frac{1}{2\pi i} \int_{-s-i\epsilon}^{-s+i\epsilon} \frac{\bar{D}( Q^2)}{Q^2} dQ^2 \ .
\end{align}
It is easy to show that to order $a^4$ the
perturbative expression for $\bar{R}_{e^+e^-}$
in terms of $\gamma_n$ and $\Pi_n$ reads:
\ea{
\bar{R}_{e^+e^-}(Q)
= &\gamma_0 + \gamma_1 a(Q) + [\gamma_2 + \beta_0 \Pi_1] a(Q)^2 
\\
& + [\gamma_3 + \beta_1 \Pi_1 + 2 \beta_0 \Pi_2- \beta_0^2 \frac{\pi^2\gamma_1}{3}  ] a(Q)^3
\nn
& + [\gamma_4 + \beta_2 \Pi_1 + 2 \beta_1 \Pi_2 + 3\beta_0 \Pi_3 
\nn
&- \frac{5}{2} \beta_0\beta_1 \frac{\pi^2\gamma_1}{3}  
- 3 \beta_0^2 \frac{\pi^2 \gamma_2}{3} -  \beta_0^3 \pi^2 \Pi_1 ] a(Q)^4  \ . 
\nonumber
}
As expected, this expression has exactly the form of Eq.~\eqref{betapattern},
with the following identification of the coefficients $r_{i,j}$:
\seal{Rdegcoef}{
r_{i,0} &= \gamma_i \\
r_{i,1} &= \Pi_{i-1} \ , \quad i\geq2 \\
r_{i,2} &= - \frac{\pi^2}{3} \gamma_{i-2} \ , \quad i\geq3\\
r_{i,3} &= -\pi^2 \Pi_{i-3} \ , \quad i\geq4
}
The expressions for the coefficients $\gamma_i$ and $\Pi_i$ can be found in Ref.~\cite{Baikov:2012zm,Baikov:2012zn}, and the four-loops $\beta$-function is given in Ref.~\cite{vanRitbergen:1997va}.
The $\gamma_i$ contain $N_f$-terms, but since they are independent of $\delta$ to any order, they are kept fixed in the scale-setting procedure.
Notice that this is a feature in dimensional regularization.

Now it is easy to set the exact PMC scales 
from Eq.~\eqref{exactscales} using that
$
R_{k,j} = (-1)^j {r_{k+j,j}}/{\gamma_k} \ ,
$
\sea{
&\ln \frac{Q_{3}^2}{Q^2}  = - \frac{\Pi_3}{\gamma_3} \ ,\\
&\ln \frac{Q_{2}^2}{Q^2}  = -\frac{\Pi_2 +\frac{1}{2}\left[\frac{\partial\beta}{\partial a}+\frac{\beta}{a}\right] \frac{\pi^2}{3}\gamma_2 }{\gamma_2-\frac{1}{2}\left[\frac{\partial\beta}{\partial a}+\frac{\beta}{a}\right]\Pi_2 } \ , \\
&\ln \frac{Q_{1}^2}{Q^2}  = \\
&\frac{- \left (\Pi_1 + \frac{1}{2}\frac{\partial\beta}{\partial a} \frac{\pi^2}{3}\gamma_1-\frac{1}{3!} \left [ \beta \frac{\partial^2 \beta}{\partial a^2} + \left( \frac{\partial\beta}{\partial a} \right )^2 \right] \pi^2\Pi_1\right )}
{\gamma_1- \frac{1}{2}\frac{\partial\beta}{\partial a} \Pi_1  +\frac{1}{3!} \left [ \beta \frac{\partial^2 \beta}{\partial a^2} - \frac{1}{2} \left( \frac{\partial\beta}{\partial a} \right )^2 \right]\frac{\Pi_1^2}{\gamma_1} - \frac{1}{4}\left(\frac{\partial\beta}{\partial a} \right)^2 \frac{\pi^2}{3} \gamma_1} .  \nonumber
}
The final resummed expression for $\bar{R}_{e^+e^-}$ reads:
\ea{
\bar{R}_{e^+e^-}(Q)
= &\gamma_0 + \gamma_1 a(Q_1) + \gamma_2 a(Q_2)^2 
\nn
 &
 + \gamma_3 a(Q_3)^3
 + \gamma_4 a(Q_4)^4 \ .
}
The scale $Q_4$ is unknown since it requires the knowledge of
the order $a^4$ coefficient of $\Pi$; to leading order it reads:
\ea{
\ln \frac{Q_{4, \rm LLO }^2}{Q^2} =  - \frac{\Pi_4}{\gamma_4} \ ,
}
however, it is possible to estimate this value. 
This is so, since $\Pi_4$ can be written as
\ea{
\label{Pi4}
\Pi _4 = &-\frac{3}{4} \beta _0^3 \Pi _{2,3}+\frac{9}{4} \beta _0^2 \Pi
   _{3,2}
   +\frac{7}{8} \beta _1 \beta _0 \Pi _{2,2}
   \nn &
   -\frac{9}{4} \beta _0 \Pi
   _{4,1}
   -\frac{1}{4} \beta _2 \Pi _{2,1}-\frac{3}{4} \beta _1 \Pi
   _{3,1}+\frac{3 \Pi _{5,0}}{4},
}
where $\Pi_{i,j}$ are the coefficients
of the \emph{bare} vacuum polarization function $\Pi_0$:
\begin{align}
\label{Pi0}
\Pi_0 (Q, a_0) & = \sum_{l=1}^\infty a_0^{l-1} \left(\frac{\mu^2}{Q^2}\right)^{l \epsilon} \sum_{k=-l}^\infty \epsilon^k \Pi_{l,k} \ ,
\end{align}
and where $\epsilon$ is the
dimensional regularization parameter, $d = 4-2 \epsilon$.
In Eq.~\eqref{Pi4} only $\Pi_{5,0}$ is unknown .
We can thus compute $Q_{4, \rm LLO}$
as a function of $\Pi_{5,0}$ and
for five active flavors we find:
\ea{
Q_{4, \rm LLO } = 0.9~e^{0.00013 \times \Pi_{5,0}}~Q \ .
 }
Because of the small partner it is reasonable to set \mbox{$Q_4 = Q$}.
The final result in numerical form in terms of $\alpha = \alpha_s/\pi$ for QCD with five active flavors reads:
\ea{
\label{Rnum}
&\bar{R}_{e^+e^-}(Q) = \frac{3}{11}  {R}_{e^+e^-}(Q) = 
\\ &
1 + \alpha(Q_1) + 1.84 \alpha(Q_2)^2
 - 1.00 \alpha(Q_3)^3 -
 11.31 \alpha(Q_4)^4 \ .
 \nonumber 
}
This is a more convergent result compared to previous estimates, and it is free of any scheme and scale ambiguities (up to strongly suppressed residual ones).

From this expression we can determine the asymptotic scale $\Lambda$ 
from the empirical data \cite{Marshall:1988ri}: 
$$\frac{3}{11} R^{\rm exp}_{e^+e^-}(\sqrt{s} = 31.6 \text{ GeV}) = 1.0527 \pm 0.0050 \ .$$  
To this end we use the logarithmic expansion for $a$ in Eq.~\eqref{alphas} and the known expressions for the $\gamma_i$ and $\Pi_i$ coefficients. For five active flavors we find:
\begin{align}
\Lambda_{\delta} =  \Lambda_{\MSB} = 419^{+222}_{-168} \text{ MeV} \ .
\end{align}
which gives us the numerical values for the PMC scales: $Q_1 = 1.3 ~Q~, Q_2 = 1.2 ~Q$, $Q_3 = 5.3 ~Q$. 
These final scales determine the effective number of quark flavors at each order of perturbation theory \cite{Brodsky:1998mf}.

Finally, the QCD coupling at the $M_Z$ scale, 
$\alpha_s(M_Z)$ can be computed using again the power expansion for $a$ in terms of $1/\ln(\mu/\Lambda)$. We find:
\begin{align}
\alpha_s(M_Z) = 0.132^{+0.010}_{-0.011} \ .
\end{align}
The error on this result is a reflection of the experimental uncertainty on
$R^{\rm exp}_{e^+e^-}$, which cannot be eliminated.
This value is somewhat larger than the present world average $\alpha_s(M_Z) =0.1184 \pm 0.0007 $, which is a global fit of all types of experiments. However, it is consistent with the values obtained from $e^+ e^-$ colliders, i.e. $\alpha_s(M_Z)=0.13\pm 0.005\pm0.03$ by the CLEO Collaboration \cite{Ammar:1997sk} and $\alpha_s(M_Z)=0.1224\pm 0.0039$ from the jet shape analysis \cite{Dissertori:2007xa}. Moreover, in computing $\alpha_s(M_Z)$ we have assumed massless quarks. The estimate will decrease when taking threshold effects properly into account\footnote{We thank Ali N. Khorramian for comments on this point.} as shown in \cite{Chetyrkin:1997sg}.

\subsection{$\tau \to \nu_\tau + $ {hadrons}}
It is straightforward to apply our results to the
$\tau$-decay into hadrons ratio
$R_{\tau} = \sigma_{\tau \to \nu_{\tau }+\text{hadrons}}/ 
\sigma_{\tau \to \nu _{\tau }+
\bar{\nu}_e+e^-}$, 
which can be computed from $R_{e^+e^-} $ \cite{Lam:1977cu}:
\begin{align}
R_{\tau} (M_\tau) =
2 \int_0^{M_\tau^2} \frac{ds}{M_\tau^2} \left(1-\frac{s}{M_\tau^2}\right)^2 \left(1+\frac{2s}{M_\tau^2}\right) \tilde{R}_{e^+e^-}(s) \ ,
\end{align}
where $\tilde{R}_{e^+e^-}$ is equal to $R_{e^+e^-}$ but with 
$\kappa = d_F \sum Q_f^2$ replaced by $\kappa' = d_F \sum | V_{ff'}|^2$, where $V_{ff'}$ are the Cabbibo-Kobayashi-Maskawa (CKM) matrix elements and $(\sum Q_f)^2 = 0$, since light-by-light scattering does not contribute.
We define in the same way $\tilde{\gamma}$Ê and $\tilde{\Pi}$ (i.e. with no light-by-light contributions).
In terms of Eq.~\eqref{betapattern},
the coefficients for the normalized $\bar{R}_\tau = R_\tau/\kappa'$ read:
\seal{taudegcoef}{
r_{i,0} &= \tilde{\gamma}_i  \ , \\
r_{i,1} &=\tilde{\Pi}_{i-1}+ \frac{19 }{12}\tilde{\gamma}_{i-1} \ , \quad i\geq2 \\
r_{i,2} &= \left (\frac{265}{72} -\frac{\pi ^2}{3}\right ) \tilde{\gamma }_{i-2} +\frac{19\tilde{\Pi }_{i-2}}{6} \ , \quad i\geq3\\
r_{i,3} &= \left ( \frac{265}{24}-\pi ^2\right )\tilde{\Pi }_{i-3}
+\left (\frac{3355}{288} -\frac{19 \pi ^2}{12} \right )\tilde{\gamma }_{i-3}
\ , \  i\geq4
}
The final expression reads
\ea{
\bar{R}_{\tau}(M_\tau)
= &\tgamma_0 + \tgamma_1 a(Q_1) + \tgamma_2 a(Q_2)^2 
\nn
 &
 + \tgamma_3 a(Q_3)^3
 + \tgamma_4 a(Q_4)^4 \ .
}
Since there are three active quark flavors for \mbox{$M_\tau \approx 1.777~ GeV$}), we find from the CKM matrix 
that $\kappa = 3(|V_{ud}|^2 + |V_{us}|^2) \approx 3$.
The effective scales read $Q_{1} = 0.67 ~Q~, Q_{2} = 0.71 ~Q$, $Q_{3} = 582 ~Q$ and for the same reason as in the case of $R_{e^+e^-}$ we set $Q_4 = Q$. 
The scale $Q_3$ has been computed to NLLO since
its $LLO$ value is smaller than the asymptotic scale $\Lambda$. 
The final result in numerical form for three active quark flavors read:
\ea{
\label{Rtaunum}
\frac{1}{3}  R_{\tau}(M_\tau) =& 1 + \alpha(Q_1) + 2.15 \alpha(Q_2)^2
\nn & + 3.44 \alpha(Q_3)^3 + 6.64 \alpha(Q_4)^4 \ ,
}
with $\alpha = \alpha_s /\pi$.
Using the asymptotic scale found from $R_{e^+e^-}$ 
we estimate the QCD contribution to the
$\tau$-decay to be:
\ea{
R_\tau(M_\tau) = 3.66^{+0.15}_{-0.22} \ .
}
This prediction is in good agreement with the
experimental result from the OPAL collaboration \cite{Ackerstaff:1998yj};
\mbox{$R_\tau^{\rm exp}(M_\tau) = 3.593 \pm 0.008$}.

\subsection{Bjorken and GLS Sum Rules}
The Bjorken sum rule \cite{Bjorken:1969mm} and the Gross-Llewellyn Smith (GLS) sum rule \cite{Gross:1969jf} obey well-known identities in conformal field theory, known as the Crewther relations \cite{Broadhurst:1993ru,Gabadadze:1995ei,Crewther:1997ux,Braun:2003rp,Baikov:2010je, Baikov:2012zn},
which through the Adler function can be used to expose the conformal terms.
In this example, we show that both sum rules after PMC scale setting
have perturbative expansions that match exactly the inverse of the anomalous dimension, $\gamma^{-1}$, and is what one expects in a conformal field theory.

The Bjorken sum rule 
expresses
the integral over the  spin distributions of quarks inside of the nucleon in terms of 
its axial charge times a  coefficient function
${C}^{Bjp}$:
\ea{
\Gamma_1^{p-n}(Q^2) &=
\int_0^1 [g_1^{ep}(x,Q^2)-g_1^{en}(x,Q^2)]dx
\nn
&=\frac{g_A}{6}
C^{Bjp}(a) +
\sum_{i=2}^{\infty}\frac{\mu_{2i}^{p-n}(Q^2)}{Q^{2i-2}}
{},
\label{gBSR}
}
where $g_1^{ep}$ and $g_1^{en}$ are the spin-dependent proton and neutron
structure functions, $g_A$ is the nucleon axial charge as measured in 
neutron $\beta$-decay. 
The sum  in the second line of Eq.~\eqref{gBSR} describes for
the nonperturbative  power corrections (higher twist) which are  inaccessible for pQCD.
Focusing on the perturbative part, we define 
\ea{
{C}^{Bjp}(Q^2) = 
1  - 3 \,C_F\, a(Q^2) +
\sum_{n=2}^{\infty} \  {\tilde{C}}_n\, a(Q^2)^n \ .
}

The Gross-Llewellyn Smith (GLS) sum rule,
\begin{equation}
\frac{1}{2}\int_0^1 F_3^{\nu p + \bar\nu p}(x,Q^2) dx = 3\, C^{GLS}(a) 
{},
\end{equation}
 relates the lowest  moment of 
the  isospin singlet structure function $F_3^{\nu p + \bar\nu p}(x,Q^2)$ to a coefficient 
$C^{CLS}(a_s)$, 
which appears in the operator product expansion of the axial and vector
non-singlet  currents. 
We are again only considering the perturbative contribution and define:
\begin{align}
{C}^{GLS}(Q^2) = 
1  - 3 \,C_F\, a(Q^2) +
\sum_{n=2}^{\infty} \  {C}_n\, a(Q^2)^n \ .
\end{align}

The 
(extended)
Crewther relation \cite{Broadhurst:1993ru,Gabadadze:1995ei,Crewther:1997ux,Braun:2003rp} states that there exists a relation 
between the two sum rules through the Adler function $D(Q^2)$ given in Eq.~\eqref{Adler} as follows:
\begin{eqnarray}
\tilde{\bar{D}}(Q^2)\, C^{Bjp}(a) &=&  1+ \frac{\beta(a)}{a}\, \tilde{K}(a) ~,
\label{gCrewtherNS}
\\
  \tilde{K}(a) &=&
a\,\tilde{K}_1 
+ a^2\,\tilde{K}_2 +a^3\,\tilde{K}_3
+ \dots  \ , 
\nonumber
\end{eqnarray}
and 
\begin{eqnarray}
{ \displaystyle \bar{D}(Q^2)\,  C^{GLS}(a)}
\label{gCrewtherFull}
&=&
 1 + 
\frac{\beta(a)}{a}\, K(a)
{},
\\
K(a) &=&
a\,K_1 
+ a^2\,K_2 +a^3\,K_3
+ \dots  \ .
\nonumber
\end{eqnarray}
The tilde on $\bar{D}$ and $K$ indicates the corresponding expressions \emph{without} the light-by-light type terms, i.e.
\sea{
\bar{D} &= \tilde{\bar{D}} + \bar{D}_{lbl}  \ , \\
K &= \tilde{K} + K_{lbl}  \ .
}
The term proportional to the $\beta$-function describes the deviation from
the limit of exact conformal invariance, with the deviations starting at order $a^2$.
Both sum rules have been explicitly computed to four loops
\cite{Bardeen:1978yd,Altarelli:1978id,Gorishnii:1985xm,Larin:1991tj,Baikov:2010je, Baikov:2012zn}
 and shown to obey the extended Crewther relations \cite{Baikov:2010je, Baikov:2012zn}%
\footnote{There is a recent claim \cite{Larin:2013yba} that the  
existing four-loop coefficient of the Bjorken sum rule \cite{Baikov:2010je, Baikov:2012zn} is missing some singlet-diagram contributions. 
This is relevant only for the explicit evaluation of $\tilde{K}_3$, 
and does not change the results of this section.}.

We can use the Crewther relations to extract the conformal and non-conformal parts of $C^{Bjp}$ and $C^{GLS}$. 
Denoting the power expansion of $\bar{D}$ by
\ea{
\label{expansion}
\bar{D}(Q^2) &= 1 +  \sum_{n=1}^\infty d_n a(Q^2)^n \ ,
}
and expanding its inverse perturbatively gives us
\ea{
&C^{GLS}(a)=
1-d_1 a+ a^2 \left[d_1^2-d_2-\beta _0 K_1\right]  \\
&+ a^3 \left[2d_1 d_2 -d_1^3-d_3+\beta _0 \left(d_1 K_1 -K_2\right) -\beta _1 K_1\right] \nn 
&+a^4 \left[d_1^4+d_2^2-d_4-3 d_1^2d_2 +2d_1 d_3+\beta _1 \left(d_1 K_1 -K_2\right)\right . \nn 
& \qquad \left . +\beta _0 \left(-d_1^2 K_1+d_1 K_2 +d_2 K_1-K_3\right) -\beta _2 K_1 \right]  \ . \nonumber
}
The expression for $C^{Bjp}$ is the same after putting tildes on the coefficients.
The $d_i$ are given in terms of $\gamma_i$, $\Pi_i$ and $\beta_i$ as follows:
\sea{
d_1 &= \gamma_1 = 3 C_F  \ , \\
d_{i\geq 2} &= \gamma_i + \sum_{k=0}^{i-2} (i-1-k) \beta_k \Pi_{i-1-k} \ .
}
We use this to find the degenerate $r_{i,j}$ coefficients of Eq.~\eqref{betapattern}.
\sea{
r_{2,1} &= - K_1 - \Pi_1 \ , \\
r_{3,1} &= - \frac{K_2}{2} - \Pi_2+ \left ( \frac{K_1}{2} + \Pi_1\right) \gamma_1  \ , \\
r_{4,1} &= -\frac{K_3}{3} -\Pi _3 +(K_2 + 4 \Pi_2)\frac{\gamma_1}{3}  
\nn
&  \qquad-\left ( \frac{K_1}{3} + \Pi_1 \right ) \gamma_1^2+( K_1+2 \Pi_1 ) \frac{\gamma_2}{3} \ ,  \\
r_{4,2} & = \frac{1}{3} ( K_1 \Pi_1 + \Pi_1^2)  \ ,  \\
r_{3,2} &= 0 \ , \quad r_{4,3} = 0  \ .
}
The degeneracy allows us to resum the series as described earlier.
The final result is:
\begin{align}
&C^{GLS}(a) =
1-a(Q_1) \gamma _1+a(Q_2)^2 \left(\gamma _1^2-\gamma _2\right)
\nn 
&+a(Q_3)^3 \left(-\gamma _1^3+2 \gamma _2
   \gamma _1-\gamma _3\right) \nn 
   &+a(Q_4)^4 \left(\gamma _1^4-3 \gamma _2 \gamma _1^2+2 \gamma _3
   \gamma _1+\gamma _2^2-\gamma _4\right)
   +{\cal O}\left(a^5\right) \ ,
\end{align}
exposing the $r_{i,0}$ coefficients.
This expression is simply the inverse of the anomalous dimension:
\begin{align}
C^{GLS}(a) = \gamma^{-1}(Q_1, Q_2, Q_3,  \ldots) \ ,
\end{align}
where we used that $\gamma_0 = 1$.
The arguments of $\gamma^{-1}$ on the right-hand side indicate
the effective scales at each order in perturbation theory, once the
inverse is Taylor expanded.
All the above expressions also apply to the Bjorken sum rules, with
the coefficients replaced by the ones with tilde.
In particular,
\mbox{
$
C^{Bjp}(a) = \tilde{\gamma}^{-1}(\tilde{Q}_1, \tilde{Q}_2, \tilde{Q}_3, \ldots)
$}.

Since, the Adler function itself after PMC scale-setting is simply given by the anomalous dimension:
\begin{align}
D(Q) = \gamma(Q_1, Q_2,Q_3,  \ldots ) \ , 
\end{align}
and correspondingly for $\tilde{D}$, the Crewther relations can be expressed as
\sea{
 \tilde{\bar{D}}(\tilde{Q}) C^{Bjp}(\mu) &= \frac{\tilde{\gamma}(\tilde{Q}_1, \tilde{Q}_2, \ldots )}{\tilde{\gamma}(\tmu_1, \tmu_2, \ldots)} = 1 \ , \\[2mm]
 {\bar{D}}(Q) C^{GLS}(\mu) &= \frac{\gamma({Q}_1, {Q}_2, \ldots )}{{\gamma}(\mu_1, \mu_2, \ldots)} = 1 \ ,
}
where the last equality follows due to conformality.
These are the generalized Crewther relations,
which set the commensurate scale relations
between the scale of the Adler function and
those of the sum rules.

\subsection{Static Quark Potential}
As a last example we consider
the potential between two static quarks, 
where the degeneracy is not explicitly apparent 
in the literature.
The static quark potential is known to order $a^4$
in the $\MSB$-scheme as an expansion in the number of massless flavors, $N_f$ \cite{Appelquist:1977tw,Peter:1996ig,Schroder:1998vy,Smirnov:2008pn,Smirnov:2009fh,Anzai:2009tm}:
\ea{
V(Q^2) = 
&- \frac{(4 \pi)^2 C_F}{Q^2} a(Q^2) 
\Big [
1 + (c_{2,0} + c_{2,1} N_f) a(Q^2) 
\nn 
&+ (c_{3,0} + c_{3,1} N_f + c_{3,2} N_f^2) a(Q^2)^2 
 \nn 
&
+(c_{4,0} + c_{4,1} N_f + c_{4,2} N_f^2+c_{4,3} N_f^3)a(Q^2)^3
\nn
& \left . 
+ 8 \pi^2 C_A^3 \ln \frac{\mu_{IR}^2}{Q^2} a(Q^2)^3 \right ] + \mathcal{O}(a^5) \ ,
}
where we have chosen the initial scale of the running coupling $\mu^2 = Q^2$,
but kept the explicit IR divergent logarithm $\ln \mu_{IR}^2/Q^2$, which is not related
to coupling constant renormalization, but is coming from
the non-Abelian gluon \mbox{`H-diagram'}.
This IR divergence is a feature of pQCD
and is canceled by non-perturbative contributions from
the 'ultra-soft' region \cite{Anzai:2009tm,Brambilla:2004jw}, which
is controlled by the domain of color-confinement.
The regularization comes from the energy difference between
color-singlet and octet intermediate states \cite{Brambilla:2004jw}.

The degenerate coefficients $r_{i,j}$ are
determined from the $c_{i,j}$-coefficients as
given in Sec.~\ref{Automation} for $n=1$.
From the explicit expression for $c_{i,j}$ given in Refs.~\cite{Smirnov:2008pn,Smirnov:2009fh,Anzai:2009tm}
we find the degenerate coefficients of the static potential to be:
\begin{subequations}
\ea{
r_{2,0}^V = &- \frac{8}{3} C_A  \label{r20} \ , \\
r_{2,1}^V = &\frac{5}{3} \ , \  
r_{3,2}^V = \left(\frac{5}{3}\right)^2 \ , \
r_{4,3}^V = \left(\frac{5}{3}\right)^3 \ , \ldots \ , 
\nonumber \\
r_{n+1,n}^V = &\left(\frac{5}{3}\right)^n \ ,  
}
\end{subequations}
\begin{widetext}
\begin{subequations}
\ea{
r_{3,0}^V = &\frac{ C_A}{36} \left[\left(532-1584 \zeta _3-9 \pi ^4+144
   \pi ^2\right) C_A +33 \left(48 \zeta _3-35\right) C_F\right] \ , \label{r30}\\
r_{3,1}^V =&\left(7 \zeta _3-\frac{217}{36}\right) C_A+\left(\frac{35}{8}-6 \zeta _3\right) C_F \ , \\
r_{4,0}^V = &\frac{11}{36} C_A \left(\left(456 \zeta _3-1440 \zeta _5+571\right) C_F^2 -9\cdot 56.83(1)
  \frac{d_F^{abcd}d_F^{abcd} }{T N_A}\right)
   +\left(-\frac{758 \zeta _3}{3}+220 \zeta
   _5+\frac{3709}{54}\right) C_A^2 C_F 
   \nn
   &+C_A^3 \left(\frac{3077 \zeta
   _3}{3}-1293.54(1)+\frac{484 \pi
   ^4}{135} \right)
   -136.39(12)\frac{d_F^{abcd}d_A^{abcd} }{N_A} \label{r40} \ ,\\
r_{4,1}^V = & \left(\frac{392 \zeta _3}{3}-20 \zeta _5-\frac{66769}{648}\right) C_A
   C_F +C_A^2 \left(196.58-192 \zeta _3-\frac{88 \pi
   ^4}{135}\right) 
   \nn
   &
   +\left(40 \zeta_5-\frac{38 \zeta _3}{3}-\frac{571}{36}\right) C_F^2+56.83(1)\frac{d_F^{abcd}d_F^{abcd} }{4 T N_A} \ , \\
r_{4,2}^V = &\left(23 \zeta _3+\frac{4 \pi ^4}{45}-\frac{2981}{144}\right)
   C_A+\left(\frac{5171}{216}-26 \zeta _3\right) C_F \ .
}
\end{subequations}
The static potential can then be written without
any explicit dependence on $N_f$:
\ea{
V(Q^2) = 
&- \frac{(4 \pi)^2 C_F}{Q^2} a(Q^2) 
\Big [
1 + (r_{2,0}^V + r_{2,1}^V \beta_0) a(Q^2) 
+ (r^V_{3,0} + \beta_1 r^V_{2,1} + 2 \beta_0 r^V_{3,1} + \beta _0^2 r^V_{3,2} ) a(Q^2)^2 
 \\ 
&
+(r^V_{4,0} + \beta_2 r^V_{2,1}  + 2\beta_1 r^V_{3,1} +  \frac{5}{2} \beta_1 \beta_0 r^V_{3,2} 
+3\beta_0 r^V_{4,1}+ 3 \beta_0^2 r^V_{4,2} + \beta_0^3 r^V_{4,3} )a(Q^2)^3
\left . 
+ 8 \pi^2 C_A^3 \ln \frac{\mu^2}{Q^2} a(Q^2)^3 \right ] + \mathcal{O}(a^5) \ .
\nonumber
}
\end{widetext}
Next, we reduce the expression to the conformal series
by using the PMC scales, which are read off
from Eq.~\eqref{effectivescales}:
\ea{
V(Q^2) = 
&- \frac{(4 \pi)^2 C_F}{Q^2} 
\Big [
a(Q_1^2) + r_{2,0}^V a(Q_2^2)^2 + r^V_{3,0}a(Q_3^2)^3
\nn  
&\left .
+r^V_{4,0}a(Q_4^2)^4
+ 8 \pi^2 C_A^3 \ln \frac{\mu^2}{Q^2} a(Q^2)^4 \right ] + \mathcal{O}(a^5) \ ,
}
The expression in the bracket defines the effective charge $a_V(Q^2)$.
Its IR divergence can be removed by adding the 
ultra-soft contributions to the static potential.
The final expression for the effective charge $\alpha_V = 4\pi a_V $ in a $SU(3)$ gauge theory with $N_f$ light quarks read:
\ea{
\alpha_V(Q^2)=
&\alpha(Q_1) -0.64 \alpha(Q_2)^2 - 0.78 \alpha(Q_3)^3 
\nn 
&+
\left(3.49 +  \frac{27}{8\pi}  \ln \frac{\mu^2}{Q^2} \right ) \alpha(Q_4^2)^4 \ ,
}
with
\sea{
Q_{1, \text{all orders}}^2 = & Q^2 \exp\left (-5/3\right ) \ ,\\
Q_{2,NLLO}^2 = & Q^2 \exp\left (0.42+0.57 \beta_0 \alpha(Q^2) \right ) \ ,\\
Q_{3,LLO}^2 = &Q^2\exp(-5.87) \ ,
}
where $\beta_0 = 11- \frac{2}{3} N_f$.
Note that the PMC scale for $Q_1$ holds to \emph{all orders}, 
which follows since 
{\mbox{$r_{n+1,n} = (5/3)^{n}$} at any order $n$.
The scales for $Q_2$ and $Q_3$ have been 
expanded in $\alpha$ to the order consistent 
with the pQCD truncation.
We note that the third PMC scale $Q_3$ is 
greatly suppressed compared to the kinematic scale $Q$.
This is to be expected physically, since
the kinematically accessible region shrinks with the loop order.
This indicates the breakdown of perturbation theory in 
higher order QCD, which is also explicitly evident by the IR divergent term appearing in the $a^4$ coefficient. 
At these higher orders non-perturbative effects must be taken into account.

Finally, we can now explicitly show that
PMC scale-setting is consistent with
the Gell Mann-Low (GM-L) scheme
and the effective coupling in QED.
It is well known that the
effective QED coupling in the massless limit,
is to leading order related to the $\overline{\rm MS}$ coupling by
a scale displacement; $\alpha_{\rm GM-L}(Q^2) = \alpha_{\overline{\rm MS}}(Q^2 e^{-5/3})$.
The effective QED coupling is precisely defined
as the effective charge of the QED static potential
between two (formally) infinitely charged particles.
By inspection of Eq.~\eqref{r20}, \eqref{r30} and \eqref{r40}
it can be seen all the higher order
conformal coefficients, $r_{i\geq 2, 0}$ are proportional
to non-Abelian group invariants, which vanish in the Abelian limit \cite{Brodsky:1997jk};
e.g. $C_A \to 0$. This generalizes to any order.
This means that the PMC expression
for the effective charge of the static potential 
is given to all orders in perturbation theory by:
\ea{
\alpha_{V, QED}(Q^2) = \alpha(Q^2 e^{-5/3}) \ .
}
Thus PMC reproduces the correct result in the Abelian limit.

\section{Commensurate Scale Relations}\label{CSR}
We demonstrate that the generic expression
in Eq.~\eqref{betapattern} extends to any scheme,
that is, the special degeneracy in the $\R_\delta$-scheme 
of an observable is inherited in all physical schemes.
This is done by relating different observables in pQCD 
using the effective charge method \cite{Grunberg:1980ja,Grunberg:1982fw,Dhar:1983py,Brodsky:1994eh}.
These commensurate scale relations 
must be independent
of the choice of scheme.
The scales are given by the systematic 
scale-setting method just described.

Any observable $\rho$ can be used to define an effective charge 
$a_\rho$.
Considering the case where the 
Born level result for the observable is just a constant
such as $R_{e^+e^-}$;
i.e. $n=0$ in Eq.~\eqref{observable}, the
effective charge is defined by the relation 
\ea{
\label{effectivecharge}
\rho (Q^2) = \rho_0(Q^2) \left [ 1 + a_\rho(Q^2) \right] \ ,
}
where $\rho_0$ is the Born (tree-level) result and
$Q^2$ is the measured scale. 
Thus, $a_\rho$ can be understood in perturbation theory as summing up the
entire perturbative series into one
effective coupling; the effective charge of the process.

It follows that the effective charge has an
expansion in the $\R_\delta$ coupling $a_\R$ 
similar to the expansion in Eq.~\eqref{betapattern}.
(we put back the index $\R$ on the coupling in this section
to avoid confusion).
By normalizing running coupling such that at leading
order the running coupling is equal to the effective charge;
i.e.
\ea{
\label{RexpansionNormalized}
a_\rho = \hat{a}_\R + \frac{r_2}{r_1^2} \hat{a}_\R^2 + \frac{r_3}{r_1^3} \hat{a}_\R^3 + \cdots \ ,
}
where $\hat{a}_\R = r_1 a_\R$,
the effective charge itself can be considered as
a running coupling of a physical scheme related to the corresponding
observable. The above expansion then defines the scheme
transformation from the $\R_\delta$-scheme to the $\rho$-scheme.
Since any two effective charges
$a_A$ and $a_B$ can be computed in
the $\R_\delta$-scheme, it follows
that $a_A$ can be written as an expansion
in $a_B$ by scheme transformations. 
Thus, any
effective charge defines a physical renormalization scheme.
The $\beta$-functions of such schemes are different
from the $\beta$-function of the $\R_\delta$-schemes,
but are related by the identity:
\ea{
\label{betaIdentity}
\beta_A (a_A) = \frac{\partial a_A}{\partial \hat{a}_\R} \beta_\R (\hat{a}_\R ) \ ,
}
where $A$ is some physical scheme corresponding to the effective charge $a_A$.
From this identity it follows that the first two coefficients of the
$\beta$-function are universal \cite{tHooft}.

The expansion of $a_A$ in $\hat{a}_\R$ can be put to the form:
\ea{
\label{effchargeA}
a_A(Q^2) &=  \hat{a}_\R(Q^2) + [r^A_{2,0} + \hat{\beta}_0 r^A_{2,1} ] \hat{a}_\R(Q^2)^2   \nn
&+[r^A_{3,0} + \hat{\beta}_1 r^A_{2,1} + 2 \hat{\beta}_0 r^A_{3,1} + \hat{\beta} _0^2 r^A_{3,2} ]\hat{a}_\R(Q^2)^3  \nn
&+[r^A_{4,0} + \hat{\beta}_2^\R r^A_{2,1}  + 2\hat{\beta}_1 r^A_{3,1} +  \frac{5}{2} \hat{\beta}_1 \hat{\beta}_0 r^A_{3,2} +3\hat{\beta}_0 r^A_{4,1}\nn
& \quad+ 3 \hat{\beta}_0^2 r^A_{4,2} + \hat{\beta}_0^3 r^A_{4,3} ] \hat{a}_\R(Q^2)^4  +{ \cal O }(\hat{a}_\R^5)  \ ,
}
where $r^A_{i,j}$ are related to the coefficients $r_{i,j}$
of the observable $A$, by:
\ea{
r^A_{i,j} = \frac{r_{i,j}}{r_1^{i-j}} \ ,
}
and $\hat{\beta}_i$ are here the coefficients of the beta function of $\hat{a}_\R$, which are related to the usual $\beta_i$ in Eq.~\eqref{beta}
by:
\ea{
\hat{\beta}_i = \frac{\beta_i}{r_1^{i+1}} \ .
}
The renormalization scheme dependence of $\hat{\beta}_2$ is denoted with a superscript.
The coefficient $\beta_2^A$ can be found in terms of $\hat{\beta}_0, \hat{\beta}_1$ and
$\hat{\beta}_2^\R$ from Eq.~\eqref{effchargeA} using Eq.~\eqref{betaIdentity}:
\ea{
\label{beta2}
\beta_2^A = &\hat{\beta}^\R _{2} -\hat{\beta} _1 r^A_{2,0}+ \hat{\beta} _0^3
   \left(r^A_{3,2}-{r^A_{2,1}}^2\right)
   \nn 
   &+2 \hat{\beta} _0^2\left (
   r^A_{3,1}- r^A_{2,0} r^A_{2,1}\right )-\hat{\beta} _0({r^A_{2,0}}^2+r^A_{3,0}) \ .
}
Using this, it can be shown that the special degeneracy in 
Eq.~\eqref{effchargeA} is preserved when relating
the effective charge $a_A$ with another effective charge $a_B$, i.e.
\ea{
\label{AB}
a_A(Q^2) &=  a_B(Q^2) + \left [r^{AB}_{2,0} + \hat{\beta}_0 r^{AB}_{2,1} \right] a_B(Q^2)^2   \\
&+\left [r^{AB}_{3,0} + \hat{\beta}_1 r^{AB}_{2,1} + 2 \hat{\beta}_0 r^{AB}_{3,1} + \hat{\beta} _0^2 r^{AB}_{3,2} \right ]a_B(Q^2)^3  \nn
&+\Big [r^{AB}_{4,0} + \beta_2^B r^{AB}_{2,1}  + 2\hat{\beta}_1 r^{AB}_{3,1} +  \frac{5}{2} \hat{\beta}_1 \hat{\beta}_0 r^{AB}_{3,2} 
\nn
& 
\quad+3\hat{\beta}_0 r^{AB}_{4,1}+ 3 \hat{\beta}_0^2 r^{AB}_{4,2} + \hat{\beta}_0^3 r^{AB}_{4,3} \Big ] a_B(Q^2)^4 \ ,
\nonumber
}
where the coefficients $r^{AB}_{i,j}$ are related to the $\R_\delta$-coefficients
as follows:
\begin{subequations}
\label{effectivechargecoefficients}
\ea{
r^{AB}_{2,0} &= r^A_{2,0}-r^B_{2,0} \ , \\
r^{AB}_{2,1} &= r^A_{2,1}-r^B_{2,1} \ ,  \\
r^{AB}_{3,0} &= r^A_{3,0}-r^B_{3,0}-2r^B_{2,0}r^{AB}_{2,0} \ ,  \\
r^{AB}_{3,1} &= r^A_{3,1}-r^B_{3,1} - r^B_{2,0}r^{AB}_{2,1}- r^B_{2,1}r^{AB}_{2,0} \ , \\
r^{AB}_{3,2} &= r^A_{3,2}-r^B_{3,2} - 2 r^B_{2,1}r^{AB}_{2,1} \ ,  \\
r^{AB}_{4,0} &= r^A_{4,0}-r^B_{4,0} - 3{r^B_{2,0}} r^{AB}_{3,0}
 - ({r^B_{2,0}}^2+2 r^B_{3,0} )r^{AB}_{2,0} \ , \\
 r^{AB}_{4,1} &= r^A_{4,1}-r^B_{4,1} - 2{r^B_{2,0}} r^{AB}_{3,1}
 - 2{r^B_{2,1}} r^{AB}_{3,0}+r^A_{3,0}r^B_{2,1} 
\nn&
 \quad -r^A_{2,1}r^B_{3,0}-
 \frac{4}{3}(r^B_{3,1}+2r^B_{2,0}r^B_{2,1})r^{AB}_{2,0}  \ , \\
  r^{AB}_{4,2} &= r^A_{4,2}-r^B_{4,2} - 2{r^B_{2,1}} r^{AB}_{3,1}
 -2r^B_{3,1}r^{AB}_{2,1} 
\nn
 &\quad - {r^B_{2,0}} r^{AB}_{3,2}  - \frac{1}{3}({r^B_{2,1}}^2+2 r^B_{3,2} )r^{AB}_{2,0} \ , \\
r^{AB}_{4,3} &= r^A_{4,3}-r^B_{4,3}  - 3{r^B_{2,1}} r^{AB}_{3,2}
 - 3 r^B_{3,2} r^{AB}_{2,0} \ .
}
\end{subequations}
This demonstrates that the special degeneracy of the
$\{\beta_i\}$ coefficients is not
a prerogative of the $\R_\delta$-schemes, but is
a general feature of perturbation theory.
Since any two effective charges are related
by the same  perturbative pattern of Eq.~\eqref{betapattern},
we can directly use the systematic scale-setting method presented
in the previous section to eliminate the initial scale ambiguity 
in Eq.~\eqref{AB}. 
This explicitly shows the renormalization-scheme invariance
of the scale-setting method to all orders in perturbation theory.
The final relation
between any two effective charges is thus:
\ea{
\label{finaleffcharges}
a_A(Q^2) =  &a_B(Q_1^2) + r^{AB}_{2,0} a_B(Q_2^2)^2  
\\ 
&
+r^{AB}_{3,0}a_B(Q_3^2)^3
+r^{AB}_{4,0}a_B(Q_4^2)^4  +{ \cal O }(a_B^5)   \ , 
\nonumber
}
where the commensurate scale relations between the two charges are exactly the PMC scales $Q_i$ as given by Eq.~\eqref{effectivescales} with
the $\beta$-function being that of $a_B$.
For completeness we provide also the expression
for the four-loop $\beta$-function coefficient of an effective charge $A$:
\ea{
\beta_3^A =& \hat{\beta} _3^{\cal R}
-2 \hat{\beta} _2^{\cal R} r^A_{2,0}
+ 
   5 \hat{\beta} _1\hat{\beta} _0^2 \left(
   r^A_{3,2}-{r^A_{2,1}}^2\right)
   \nn
   &+
   4\hat{\beta} _1\hat{\beta} _0 \left( r^A_{3,1}- r^A_{2,0}
   r^A_{2,1}\right)+\hat{\beta} _1 {r^A_{2,0}}^2
\nn
&+
2\hat{\beta} _0^4 \left(2 {r^A_{2,1}}^3-3 r^A_{3,2} r^A_{2,1}+ r^A_{4,3}\right)
\nn
&+
6\hat{\beta} _0^3
   \left(2 r^A_{2,0} {r^A_{2,1}}^2-2 r^A_{3,1} r^A_{2,1}- r^A_{2,0} r^A_{3,2}+
   r^A_{4,2}\right)
   \nn
   &+
   6\hat{\beta} _0^2 \left( 2r^A_{2,1} {r^A_{2,0}}^2-2 r^A_{3,1} r^A_{2,0}-
   r^A_{2,1} r^A_{3,0}+ r^A_{4,1}\right)
   \nn
   &+
   2\hat{\beta} _0 \left(2 {r^A_{2,0}}^3-3 r^A_{3,0}
   r^A_{2,0}+ r^A_{4,0}\right) \ .
}

As a particularly simple example, 
we relate the effective charge of $R_\tau$; $a_\tau$, to
that of $R_{e^+e^-}$; $a_R$ and apply the
systematic scale-setting 
method to derive commensurate scale relations between
the two effective charges. 
The final result is completely independent of 
the intermediate renormalization scheme and scale
used to compute $a_\tau$ and $a_R$.

The degenerate coefficients of $a_\tau$ and $a_R$ in
the $\R_\delta$-scheme can be read off from
Eq.~\eqref{taudegcoef} and \eqref{Rdegcoef},
from which we compute $r^{\tau, R}_{i,j}$.
Using Eq.~\eqref{effectivechargecoefficients} 
and Eq.~\eqref{finaleffcharges} we can readily
express $a_\tau$ as a perturbative series in $a_R$:
 \ea{
 a_\tau(Q^2) = &a_R(Q_{R,1}^2) - \frac{\gamma_{3, lbl}}{\gamma_1^3} a_R(Q_{R,3}^2)^3
\nn &
-\frac{\gamma_1\gamma_{4, lbl}-3\gamma_2 \gamma_{3, lbl} }{\gamma_1^5} a_R(Q_{R, 4}^2)^4 \ ,
}
where the PMC scales $Q_{R, i}$ are given by
the systematic method and where $\gamma_{i,lbl}$ is the
light-by-light part of $\gamma_i$; i.e. the
two effective charges are equivalent up to light-by-light terms.
The corrections start at order $a^3$, since there is no light-by-light contribution to $\gamma_2$.
The PMC scales expanded in $a_\tau(Q)$ read:
\sea{
\ln \frac{Q_{R,1}^2}{Q^2} = &-\frac{19}{12} -\frac{169}{144} \hat{\beta} _0 a_R(Q^2 )
\nn
&-
\left(\frac{761 \hat{\beta} _0^2}{192}+\frac{169 \hat{\beta} _1}{96}\right) a_R(Q^2
   )^2  \ , \\
  \ln \frac{Q_{R,3}^2}{Q^2} = & 
  \left (\frac{19}{12}
  +\frac{\Pi _1 }{\gamma _1}\right) \frac{\gamma _{3,lbl}}{{\gamma}_3}-\frac{\Pi _{3,lbl}}{{\gamma}_3} \ .
 }
 
By definition the scale where the expression for the effective charge $a_\tau$
applies is $Q^2 = M_\tau^2$.  At this scale, 
the number of light flavors is $N_f= 3$.
Light-by-light diagrams are proportional to $(\sum_f Q_f)^2$, which
vanishes exactly when summing over the three light quarks.
Therefore in three-flavor QCD the two effective charges are identical to all orders; i.e.
using 
\mbox{$a_{R/\tau} = \gamma_1 \alpha_{R/\tau}/(4\pi) = \alpha_{R/\tau}/\pi$},
we have
\ea{
\frac{\alpha_\tau(M_\tau^2)}{\pi} = \frac{\alpha_R(Q_{R,1}^2)}{\pi} \ ,
}
where 
the commensurate scale for $a_R$ up to four-loop order is given by:
\ea{
\ln \frac{Q_{R,1}^2}{M_\tau^2} = &-\frac{19}{12} -\frac{169}{64}  \frac{\alpha_R(M_\tau^2 )}{\pi}- \frac{83273}{3072}\frac{\alpha_R(M_\tau^2)^2}{\pi^2} \ .
}
This relation (at one lower order) has been shown to be in very good agreement with experiment~\cite{Brodsky:2002nb} demonstrating a highly
nontrivial consistency check of QCD, free of any scheme and scale ambiguities.

\section{Conclusion}
In this paper we have shown that a generalization of the conventional $\MSB$-scheme is illuminating. It enables one to determine the general (and degenerate) pattern of nonconformal $\{\beta_i\}$-terms and to systematically determine the argument of the running coupling order by order in pQCD, in a way which is readily automatized. The resummed series matches the conformal series, in which no factorially divergent $n! \beta^n \alpha_s^n$ ``renormalon" series appear 
and which is free of any scheme and scale ambiguities.
Thus using the PMC/BLM procedure, all non-conformal contributions in the perturbative expansion series are summed into the running coupling by shifting the renormalization scale in $\alpha_s$ from its initial value, and one obtains unique, scale-fixed, scheme-independent predictions at any finite order.  The resulting PMC scales and finite-order PMC predictions are both to high accuracy independent of the choice of initial renormalization scale. The PMC procedure also provides scale-fixed, scheme-independent commensurate scale relations, relations between observables which are based on the underlying conformal behavior of QCD such as the generalized Crewther relation.
Furthermore, we have shown that PMC is consistent with
QED scale-setting, where there is no ambiguity in choosing the final scale of the effective coupling.
The PMC satisfies all of the principles of the renormalization group: reflectivity, symmetry, and transitivity, and it thus eliminates an unnecessary source of systematic error in pQCD predictions.

\vfill
\begin{acknowledgments}
We thank Konstantin Chetyrkin, Joseph Day, Leonardo Di Giustino, Stefan H\" oche, Andrei L. Kataev and Ali N. Khorramian for useful comments.
We are grateful to Sheng-Quan Wang for carefully proof-reading the manuscript.
MM thanks SLAC theory group for kind hospitality. This work was supported in part by the Department of Energy contract DE-AC02-76SF00515, the Natural Science Foundation of China under Grant NO.11275280 and the Danish National Research Foundation, grant no. DNRF90.
\end{acknowledgments}

\appendix

\section{
On the discrepancy between the new PMC results
and previous BLM based results
}

In this Appendix we address the question raised by A. L. Kataev in 
\cite{Kataev:2013vua}%
\footnote{
We note that in the second version of this paper, the criticism on our work was withdrawn. Nevertheless, this Appendix might still be helpful to avoid future confusion.}
about
the discrepancy of the new PMC result
for the Adler function in this work and Refs. \cite{Mojaza:2012mf,Wu:2013ei} and previous results based on BLM.
Let us immediately stress that the discrepancy occurs due to the still
unsettled question of which $n_f$ terms in the perturbative
coefficients should be treated as conformal terms and which
should be related to renormalization (non-conformal terms).
This statement applies to perturbative QCD in general and therefore also applies to the PMC results given in this work for $R_{e^+e^-}$, $R_{\tau \to \nu + \textbf{h}}$, the Bjorken sum rule and the GLS sum rule.

Let us first remind that the purpose of BLM/PMC scale-setting 
is to set the scales of perturbative QCD in such a way that
the scheme and scale ambiguities of the final expression are
essentially eliminated. 
This goal can be reached by understanding the separation 
of the conformal and non-conformal contributions to the perturbative series.

{Note that the definition we are using for the conformal series is the one in which all $\{\beta_i\}=0$.}
The definition used in \cite{Kataev:2013vua} seems to be different.
On the other hand the same definition is used in some of the previous BLM literature, so why is there a discrepancy?
The reason is that in the previous literature the explicit $n_f$-series
has been used to identify the non-conformal terms, while
in our recent papers we have used the $\R_\delta$-scheme.
We believe that the latter is more correct for the following reason:
{Using the explicit $n_f$-series, one is forced to relate \emph{all} $n_f$-terms to renormalization of the coupling which is strictly incorrect, but a good approximation, at least to the orders known in pQCD.
Instead, one should treat the $n_f$-terms unrelated to renormalization of the coupling as part of the conformal coefficient}; e.g.,
the $n_f$-terms coming from light-by-light scattering in QED and the $n_f$-terms unrelated to the renormalization of the tri-gluon and quartic-four-gluon vertices belongs to the conformal series 
(see e.g. \cite{Binger:2006sj}).

Using the $\R_\delta$-scheme one can instead derive the
$\{\beta_i\}$-series expansion of the perturbative coefficients,
thereby avoiding the use of the $n_f$-series expansion.
So far no known pQCD results have been
computed in the $\R_\delta$-scheme. 
However, for the particular cases of
the $R_{e^+e^-}$, $R_{\tau \to \nu + \textbf{h}}$, the Bjorken sum rule and the GLS sum rule we have been able to
derive the $\{\beta_i\}$-series expansion. 
This is possible because they are all related to the Adler function,
which can be explicitly written as a sum of conformal and non-conformal contributions:
\ea{
\label{Adler}
D(Q^2) = \gamma(a) + \beta(a) \frac{d}{da} \Pi(Q^2,a)
}
Thus, we have suggested that the PMC result for
the Adler function reads:
\ea{
D(Q^2) = \gamma_0 + \gamma_1 \alpha(Q_1) + \gamma_2\alpha(Q_2)^2 + \gamma_3 \alpha(Q_3)^3 + \cdots \ ,
}
which in numerical form, known up to four loops \cite{Baikov:2012zm}, reads (we neglect for simplicity the light-by-light or singlet type terms):
\ea{
D(Q^2) = &\sum_{f=1}^{n_f} Q_f^2 \Big [ 1+ \alpha(Q_1) + (2.60 - 0.15 n_f)\alpha(Q_2)^2
\nn
 &
 + \left (9.74 -2.04 n_f -0.02 n_f^2 \right ) \alpha (Q_3)^3  
 \nn 
 &
 + \left ( 41.09 - 13.00 n_f - 0.49 n_f^2 + 0.005 n_f^3 \right ) \alpha (Q_4)^4  \Big ] 
\nn
&
 + \mathcal{O}(\alpha^5)
}
where $\alpha = \alpha_s/\pi$ and the PMC scales $Q_i$ are functions of $Q$ and
are derived as given in Sec.~\ref{Systematic}. Their values are unimportant for the purpose of this Appendix.
{Notice that $n_f$ terms enter already from NLO and beyond.
We propose that these $n_f$ terms should not be absorbed
into the running coupling since 
they are independent of which renormalization scheme is used in dimensional regularization.}
The appearance of these $n_f$-terms is the reason behind the discrepancy. 

On the other hand, as also described in Sec.~\ref{Automation},
when one is confronted with not knowing the 
$\{\beta_i\}$-series expansion, the explicit $n_f$-series
can be used as an approximation to get the PMC result.

Suppose we did not have Eq.~\eqref{Adler} in hand and
instead only knew the Adler function in perturbation theory
with its $n_f$-series for each coefficient.
This reads:
\ea{
D(Q^2) = &\sum_{f=1}^{n_f} Q_f^2 \Big [ 1+ \alpha(Q)
+ (1.99-0.12 n_f) \alpha(Q)^2 
\nn
&
+\left(18.24-4.22 n_f+ 0.086 n_f^2 \right)\alpha(Q)^3
\nn
&
+ \left(135.87 -34.52 n_f+1.88 n_f^2+ 0.01 n_f^3\right)\alpha(Q)^4 \Big ] \ ,
\nn
&
+O\left(\alpha ^5\right)
 }
 where we have set the initial renormalization scale $\mu = Q$.
 From Sec.~\ref{Automation} we then find the PMC result to be:
\ea{
D(Q^2) = &\sum_{f=1}^{n_f} Q_f^2 \Big [ 1+ \alpha(\hat{Q}_1)
+ 0.083 \alpha(\hat{Q}_2)^2 
\nn
&
-23.22 \alpha(\hat{Q}_3)^3
+ 81.24 \alpha(\hat{Q}_4)^4 \Big ]+O\left(\alpha ^5\right)
}
{Notice that up to NLO this, as expected, agrees with
all the previous BLM literature including 
the recent paper of A. L. Kataev \cite{Kataev:2013vua}}.
For the higher order terms a careful treatment of the 
light-by-light terms must be made.
We stress again that in this result
\emph{the effective PMC scales $\hat{Q}_i$ include
absorption of $n_f$-terms that are unrelated to renormalizing
the running coupling}, and the expression is therefore only
an approximation.

For $R_{e^+e^-}$ both methods were used in 
Ref.~\cite{Wu:2013ei}, where it was found that the final numerical results of the two methods are consistent with each other. Furthermore, a recent analysis on the Higgs boson inclusive decay channels 
$H\to b\bar{b}$ and $H\to gg$ up to four loops \cite{Wang:2013bla} shows that also here the two treatments are consistent with each other and even with the BLM scale-setting approach proposed in Ref.~\cite{Mikhailov:2004iq}, which also resorts to the explicit use of the $n_f$-series.
Thus, the
explicit $n_f$-series treatment seems to be a good approximation for phenomenological applications.
\\

This explicit treatment should explain
the discrepancies between the conformal coefficients
in this work and the previous ones 
\cite{Mojaza:2012mf,Wu:2013ei} (based on the $\cal R_\delta$-scheme)
and those found previously in the BLM literature (based on the $n_f$-series expansion).
Here, we have explained why we believe the former treatment is
formally more correct, while in some applications the results are effectively the same.

\end{document}